\documentclass[a4paper]{article}
\usepackage{jheppub}
\usepackage{dsfont}
\usepackage{amsmath,amssymb,amscd,amsfonts,mathtools}

\usepackage[toc,page]{appendix}
\usepackage{epsfig}
\usepackage{epstopdf}
\usepackage{latexsym}
\usepackage{graphicx}
\usepackage{placeins}
\usepackage{floatrow}
\usepackage{subfig}
\usepackage{caption}
\usepackage{booktabs}
\usepackage{bbm}
\usepackage{color}
\usepackage{physics}
\usepackage{tensor}
\usepackage{tikz}
\usepackage{comment}
\usepackage{accents}
\usetikzlibrary{matrix}
\usetikzlibrary{decorations.markings,calc,shapes,decorations.pathmorphing,patterns,decorations.pathreplacing,arrows.meta}
\usetikzlibrary{positioning}
\usepackage{hyperref}

\pdfoutput=1
\makeatletter
\def\@fpheader{\relax}
\makeatother

\newcommand*{\rh}[1]{%
  \accentset{\mbox{\scriptsize\bfseries .}}{#1}}
\newcommand*{\lh}[1]{%
  \accentset{\mbox{\phantom{\scriptsize\bfseries .}}}{#1}}

\title{Gravito-electromagnetism, Kerr-Schild and Weyl double copies; a unified perspective}

\author{Elena C\'aceres$^*$}
\author{Brian Kent$^*$}
\author{Harita Palani Balaji$^*$}
\affiliation{$^*$Theory Group, Weinberg Institute, Department of Physics, University of Texas, 2515 Speedway, Austin, Texas 78712, USA.}

\abstract{Two modern programs involving analogies between general relativity and electromagnetism, gravito-electromagnetism (GEM) and the classical double copy (CDC), induce electromagnetic potentials from specific classes of spacetime metrics. We demonstrate such electromagnetic potentials are typically gauge equivalent to Killing vectors present in the spacetime, long known themselves to be analogous to electromagnetic potentials. We utilize this perspective to relate the Type D Weyl double copy to the Kerr-Schild double copy without appealing to specific coordinates. We analyze the typical assumptions taken within Kerr-Schild double copies, emphasizing the role Killing vectors play in the construction. The basis of the GEM program utilizes comparisons of tidal tensors between GR and EM; we perform a more detailed analysis of conditions necessary for equivalent tidal tensors between the theories, and note they require the same source prescription as the classical double copy. We discuss how these Killing vector potentials relate to the Weyl double copy, in particular there must a relation between the field strength formed from the Killing vector and the Weyl tensor. We consider spacetimes admitting a Killing-Yano tensor which provide a particularly insightful example of this correspondence. This includes a broad class of spacetimes, and provides an explanation for observations regarding the splitting of the Weyl tensor noted when including sources.}

\begin{document}	
\maketitle
\flushbottom
\vfill\pagebreak

\section{Introduction}

Analogies between general relativity (GR) and electromagnetism (EM) have a lengthy history. Generically they have clear differences in behavior, nevertheless there are situations where the mathematics of electromagnetism have drawn analogy to, and sometimes even been useful for analyzing structures within general relativity. A small sampling of such examples include Geroch-Hansen multipole moments \cite{Geroch1970-yf,Hansen1974-xa}, where vacuum stationary and asymtotically flat spacetimes have a structure built from Killing vector fields matching a multipole expansion. The Bel-Robinson tensor \cite{Bel} can be viewed as a rank four analogy to the electromagnetic stress energy tensor instead constituted by the Weyl tensor, and having further analogy in vacuum from being divergence free and satisfying similar energy conditions \cite{Penrose_Rindler_1984}. As a final example relevant for this work, the mathematical structure of Killing vector fields in vacuum spacetimes mimics that of a vacuum electromagnetic field, and has been utilized for considering \emph{test} (non-spacetime sourcing) electromagnetic fields in general relativity \cite{Papapetrou,Wald1974b}. We seek to understand how two modern programs of \emph{exact analogies}, in the sense of identical structure between a test electromagnetic field and gravitational field, are related. Specifically, the gravito-electromagnetism and the classical double copy programs.

The \emph{classical double copy} finds its origins through quantum field theory scattering amplitudes, where it was demonstrated \cite{Bern:2008qj} that gauge theory amplitudes could be split into ``color" and ``kinematic" pieces both obeying the same algebra, known as \emph{color-kinematics} duality. The construction of new amplitudes via replacing color and kinematic factors, and the program itself, goes by the name of the \emph{double copy}. Such a relation is also a low-energy prediction of tree-level relations in string theory between open and closed strings, known as the KLT relations \cite{Kawai:1985xq}. Perhaps the most well known construction relates gluon amplitudes to graviton amplitudes (along with a dilaton and 2-form), with the gluon amplitudes themselves constructable from a bi-adjoint scalar theory.

That such relations have continued to hold (although not generally proven) for higher loop amplitudes \cite{Bern:2013yya} suggests that analogous relations might hold at the level of Einstein's equations. The first such proposal \cite{Monteiro:2014cda} utilized \emph{Kerr-Schild} metrics \cite{1965Kerr}:
\begin{equation}
    g_{\mu \nu} = \eta_{\mu \nu} + \psi k_\mu k_\nu.
\end{equation}
Under specific assumptions (enumerated later), the ``graviton" $h_{\mu \nu} = \psi k_\mu k_\nu$ can have ``kinematic factors" $k_\mu$ removed to produce other theories. Namely $A_\mu = \psi k_\mu$ satisfies Maxwell's equations on $g_{\mu \nu}$\footnote{As a ``test" electromagnetic field, meaning it does not source the spacetime corresponding to $g_{\mu \nu}$.} or Minkowski space $\eta_{\mu \nu}$, known as the \emph{single-copy}, and the scalar function $\psi$ satisfies the abelianized bi-adjoint scalar equations of motion on $\eta_{\mu \nu}$, $\Box\psi=0$, known as the \textit{zeroth copy}.

The electromagnetic theory, known as the ``single-copy", often has an imprint of the features present from the original spacetime. For instance, a Schwarzschild black hole is analogous to a point charge, while a Kerr black hole is analogous to a rotating or ring charge \cite{Monteiro:2014cda}, and the Taub-NUT spacetime is like a magnetic monopole \cite{Luna:2015paa}, with even the gravitational horizons being analyzed \cite{Chawla:2023bsu,He:2023iew}. Both plane waves and shockwaves have had correspondences drawn \cite{Monteiro:2014cda,Bahjat-Abbas:2020cyb}, so too have gravitational waves \cite{Godazgar:2020zbv,CarrilloGonzalez:2022mxx,Andrzejewski:2019hub,Ilderton:2018lsf} as well as electromagnetic duality manifesting as an Ehlers transform in general relativity \cite{alawadhi2020s}. There are countless other examples, which serve to demonstrate that there exists a rich array of correspondences. There have also been several investigations into understanding the fundamental structure of classical double copies. At linearized level, twistor methods have connected the spinorial version of the classical double copy \cite{Luna:2018dpt} to momentum space double copy amplitudes \cite{Chacon:2021wbr,Chacon:2021lox,Luna:2022dxo}. An extension covering all vacuum Kerr-Schild spacetimes was considered in \cite{Easson:2023dbk}, and curved space generalizations have been considered in \cite{Carrillo-Gonzalez:2017iyj,Bahjat-Abbas:2017htu,Alka__2021,Alkac:2021bav}.

On the other hand, the \textit{gravito-electromagnetism} (GEM) program \cite{Costa_2008,Costa2014} concerns itself with comparing the tidal tensor structures between gravity and electromagnetism. The methodology involves considering the relative acceleration of nearby particles with initially identical velocity undergoing the same force (implying same charge-to-mass ratio for charged particles), which can be mathematically described through tidal tensors. These are defined by the geodesic deviation equation for general relativity, and within electromagnetism an electric tidal tensor $E_{\mu \nu}$ dependent on some observer. Furthermore, magnetic tidal tensors can be defined in both theories, representing tidal forces on extended spinning bodies in GR, and a force on a magnetic dipole in the case of EM. That these tidal tensors are generically different should come as no surprise, \cite{Costa2014} does an excellent job contrasting the symmetries and physical interpretation of such structures between the two theories. Nevertheless, several classes of stationary metrics have been considered at both linearized and exact solution level \cite{Costa_2008} which contain structures analogous to an electromagnetic potential in their metrics which exhibit identical tidal tensors for the gravitational and electromagnetic theories.

An alternative perspective to the Kerr-Schild double copy was provided utilizing the spinorial formulation of gravity, known as the \textit{Weyl Double Copy} \cite{Luna:2018dpt}. Based upon the existence of test electromagnetic fields within Type D vacuum spacetimes noted in \cite{Walker:1970un,Hughston:1972qf}, the Weyl Double Copy concerns the splitting of the Weyl spinor (satisfying the massless spin-two free-field equation in vacuum) into electromagnetic spinors (satisfying the massless spin-one free-field equation in vacuum). This procedure was generalized to certain sourced spacetimes (including black holes with electric or magnetic charge) in \cite{Easson:2021asd,Easson:2022zoh,alkac2024regularizedweyldoublecopy}. In those works, they noted a characteristic splitting of the electromagnetic spinors into a vacuum piece and pieces sourced (in the sense of Maxwell's equations) by the Ricci tensor. While this has been proven at linearized level utilizing twistors \cite{Armstrong-Williams:2024bog}, it remains unclear how such a splitting arises at non-linear level. More generally these constructions have relied on specific coordinatizations to prove equivalence between the constructions, although at least one example of a Type D sourced Weyl double copy exists \cite{Kent2025} which does not have a Kerr-Schild representation.

We demonstrate these programs typically utilize an electromagnetic potential gauge equivalent to a Killing vector present in the spacetime, and that this new perspective sheds light on several previously unexplained features. Specifically, electromagnetic tidal tensors formed from a Killing vector necessarily satisfy exact tidal analogies with gravitational tidal tensors. These tidal tensors rely on an observer, which are only physically interpretable if the Killing vector is itself geodesic. Meanwhile for the Weyl double copy, we investigate spacetimes admitting a Killing-Yano tensor, which include a broad class of studied examples in the literature. These provide a particularly clean example of how field strengths built from a specific Killing vector is directly related to the Weyl tensor. Some spacetimes analyzed in the literature outside this class include magnetic charges on the single-copy side, and so cannot have an equivalence with Killing vector potentials. However spacetimes outside this class that have an equivalence with the Kerr-Schild double copy may have some understanding in terms of Killing vectors. We provide some discussion on how such equivalences may arise.

We emphasize there is more to these programs than just the reliance on a Killing vector. In the classical double copy, metrics considered allow the electromagnetic theory to satisfy Maxwell's equations on both curved and flat space, which is far more constraining than simply admitting a Killing vector within the spacetime. In the case of gravito-electromagnetism, the Killing vector must be a timelike geodesic to represent an observer, or for tidal tensors to even be applicable. Nevertheless, for both programs many constructions themselves still rely on subclasses of spacetimes with Killing vectors from which an electromagnetic field strength is constructed, which is the connection of the two programs.

This paper is organized as follows: in section \ref{sec2}, we introduce the Kerr-Schild double copy, as well as the properties of Killing vectors treated as electromagnetic potentials. We demonstrate that the typical potentials within the Kerr-Schild double copy are gauge equivalent to a Killing vector. We analyze conditions for an electromagnetic field built from a Killing vector to satisfy equivalent Maxwell's equations on the full spacetime and base metric, as is true in the Kerr-Schild double copy. In section \ref{sec3} we provide an introduction to, and derivations for, tidal tensors in general relativity including an electromagnetic field. The general conditions for an equivalence between electromagnetic and gravitational tidal tensors are established, demonstrating that they require the same source correspondence as the Kerr-Schild double copy. In section \ref{sec4} we demonstrate that taking Killing vectors as an electromagnetic potential automatically satisfies exact tidal analogies. Furthermore, we show that the typical examples presented in the literature are gauge equivalent to a Killing vector potential, hence connecting the gravito-electromagnetism and the Kerr-Schild double copy programs. We conclude this section with comments and discussion on the physical interpretations made within such analogies. We begin section \ref{sec5} with a brief introduction to the spinor formalism in general relativity and how it is utilized within the Weyl double copy. Properties of Killing-Yano tensors are discussed, which allow a direct correspondence between a field strength built from a Killing vector and the Weyl tensor. Known prominent examples are demonstrated to admit Killing-Yano tensors, and those which do not are subsequently addressed.

\section{Reframing Kerr-Schild double copies}\label{sec2}

Before moving on to calculations, the conventions utilized in this work are set, namely signature $(-+++)$, natural units $c = G = 1$, with Riemann and Maxwell defined as:
\begin{equation}
    \nabla_\nu F^{\mu \nu} = 4 \pi J^\mu, \quad 2 \nabla_{[\mu}\nabla_{\nu]}\omega_\beta = -\omega_\alpha \tensor{R}{^\alpha_{\beta \mu \nu}}.
\end{equation}

\subsection{The Kerr-Schild double copy}

The original construction of the Kerr-Schild double copy utilizes metrics of the form:
\begin{equation}
    g_{\mu \nu} = \eta_{\mu \nu} + \psi k_\mu k_\nu, \label{KSmetric}
\end{equation}
where $\eta_{\mu \nu}$ is a flat base metric, $\psi = \psi(x^\mu)$ is a coordinate dependent function, and $k_\mu$ is a null vector. A straightforward calculation demonstrates that the null vector $k_\mu$ may be raised with respect to either the base metric or the full metric. 

As has been emphasized in the context of classical double copies in \cite{Easson:2023dbk}, Einstein's vacuum equations imply that $k_\mu$ is geodesic. For sourced spacetimes this is typically taken as an assumption, as many nice results follow (see \cite{Stephani:2003tm}). For instance, the null vector $k_\mu$ is geodesic if and only if $k_\mu$ is an eigenvector of the stress-energy tensor, also if and only if it is a multiple principle null vector of the Weyl tensor (and so is algebraically special in the sense of the Petrov classification). Of importance here, $k_\mu$ being geodesic implies the mixed index Ricci tensor is linear in the function $\psi$: 
\begin{equation}
     \tensor{R}{^\mu_\nu} = \frac{1}{2}\eta^{\mu \alpha}\eta^{\beta \gamma}\nabla^{(0)}_{\beta}\big[\nabla^{(0)}_{\alpha}(\psi k_\gamma k_\nu) + \nabla^{(0)}_{\nu}(\psi k_\gamma k_\alpha) - \nabla^{(0)}_{\gamma}(\psi k_\alpha k_\nu) \big], \label{linearKS}
\end{equation}
where $\nabla^{(0)}_\mu$ is the covariant derivative with respect to $\eta_{\mu \nu}$. From here the existence of a stationary Killing vector $\xi^\mu$ is typically assumed, which is covariantly constant with respect to $\eta_{\mu \nu}$, meaning $\nabla^{(0)}_\mu \xi^\nu = 0$. Choosing coordinates such that the metric is independent of $x^0$, and choosing $k_\mu \xi^\mu = k_0 = 1$ yields:
\begin{equation}
     2\tensor{R}{^\mu_\nu}\xi^\nu = \nabla^{(0)}_{\nu}\left[\nabla_{(0)}^{\mu}(\psi k^\nu) -\nabla_{(0)}^{\nu}(\psi k^\mu) \right] \equiv \nabla^{(0)}_{\nu}F^{\mu \nu}, \label{Fddflat}
\end{equation}
where raised coordinates on the right-hand side are with respect to the base metric $\eta^{\mu \nu}$. This equation also implies a sourced wave equation $\nabla^{(0)}_\mu \nabla_{(0)}^\mu \psi = -2 R_{\mu \nu}\xi^\mu \xi^\nu$, known as the zeroth copy. Note that the flat space Maxwell equation is really just a simplification of the full curved space case:
\begin{equation}
    2\tensor{R}{^\mu_\nu}\xi^\nu = \nabla_{\nu}\left[\nabla^{\mu}(\psi k^\nu) -\nabla^{\nu}(\psi k^\mu) \right] \equiv \nabla_{\nu}F^{\mu \nu}, \label{Fddfull}
\end{equation}
and so it is equally valid to consider this as a solution to Maxwell's equation on the original curved spacetime, see \cite{Kent2025} for details. However, the zeroth copy fails to hold in this case, meaning $\nabla_\mu \nabla^\mu \psi \neq -2 R_{\mu \nu}\xi^\mu \xi^\nu$. The primary result is that given the above assumptions, the one-form $A_\mu \equiv \psi k_\mu = \psi k_\mu (k_\nu \xi^\nu)$ (equivalent due to normalization of $k_0$) acts as an electromagnetic potential sourced by the stress-energy of the spacetime through Einstein's equations. There were several assumptions necessary for such a construction to work, and so which assumptions were responsible for each property is generically difficult to disentangle. In the following sections, we seek to reframe these assumptions, beginning with demonstrating typical potentials are gauge equivalent to Killing vectors.

\subsection{Killing vector potentials and gauge equivalence}

As first noted in \cite{Papapetrou} and expanded upon in \cite{Wald1974b,Wald:1984rg}, consider a Killing vector $\xi^\mu$:
\begin{equation}
    \nabla_{(\mu}\xi_{\nu)} = 0,
\end{equation}
and a two-form field defined as:
\begin{equation}
    F_{\mu \nu} = 2\nabla_{[\mu}\xi_{\nu]} = 2\nabla_\mu \xi_\nu.
\end{equation}
Its relation to the Riemann tensor can be determined through:
\begin{equation}
    \left(\nabla_\beta \nabla_\alpha - \nabla_\alpha \nabla_\beta \right)\xi_\mu = \xi_\nu \tensor{R}{^\nu_{\mu \alpha \beta}} = \nabla_\beta \nabla_\alpha \xi_\mu + \nabla_\alpha \nabla_\mu \xi_\beta. \label{FtoR}
\end{equation}
Adding the permutation $(\mu \alpha \beta) \to (\beta \mu \alpha)$ and subtracting the permutation $(\mu \alpha \beta) \to (\alpha \beta \mu)$ yields the formula:
\begin{equation}
    \nabla_\mu F_{\alpha \beta} = 2\nabla_\mu \nabla_\alpha \xi_\beta = 2\xi_\nu\tensor{R}{^\nu_{\mu \alpha \beta}}. \label{2frmK}
\end{equation}
This two-form therefore immediately satisfies the Bianchi identity:
\begin{equation}
    \nabla_{[\mu}F_{\alpha \beta]} = 2\xi_\nu\tensor{R}{^\nu_{[\mu \alpha \beta]}} = 0, \label{bianchi}
\end{equation}
and if treated as a Maxwell field, is sourced by the Ricci tensor as:
\begin{equation}
    \nabla_\nu F^{\mu \nu} = 4\pi J^\mu = 2\tensor{R}{^\mu_\nu}\xi^\nu. \label{KillingSource}
\end{equation}
One may wonder what the implication of $\nabla_\mu J^\mu = 0$ is with respect to the Killing vector, and it implies the known fact that $\xi^\mu \nabla_\mu R = 0$ which follows from $\mathcal{L}_\xi \tensor{R}{^\mu_{\nu \alpha \beta}} = 0$ and $\mathcal{L}_\xi g_{\mu \nu} = 0$, where $\mathcal{L}_\xi$ is the Lie derivative along $\xi^\mu$. Note this is the exact same source prescription as \eqref{Fddflat}. Now consider a metric splitting of the form:
\begin{equation}
    g_{\mu \nu} = \hat{g}_{\mu \nu} + h_{\mu \nu},
\end{equation}
where $\hat{g}_{\mu \nu}$ is also a solution to Einstein's equations. If $\xi^\mu$ is a Killing vector on the metric $g_{\mu \nu}$, and is furthermore an exact form on $\hat{g}_{\mu \nu}$ then:
\begin{equation}
    \xi_\mu = \hat{g}_{\mu \nu}\xi^\nu + h_{\mu \nu}\xi^\nu \equiv \partial_\mu \lambda + A_\mu,
\end{equation}
for some scalar function $\lambda$, and we have defined $A_\mu \equiv h_{\mu \nu}\xi^\nu$. These assumptions make evident that $\xi_\mu$ and $A_\mu$ are gauge equivalent when constructing $F_{\mu \nu}$. In the case of vacuum and electro-vacuum Kerr-Schild metrics \eqref{KSmetric}, it is known that a Killing vector which is exact on $\eta_{\mu \nu}$ always exists \cite{1969JMP....10.1842D}. In the context of vacuum Kerr-Schild double copies, this observation was utilized in \cite{Easson:2023dbk} to generalize beyond stationary Killing vectors. For specific examples, this correspondence was pointed out in \cite{Ortaggio_2024}, and furthermore at least one class of spacetimes with a single copy field, general Type N vacuum Kundt metrics for which vacuum pp-waves form a Kerr-Schild subclass, are not generally constructable via a Killing vector \cite{Ortaggio_2024b}. It is worth noting however that for type N vacuum spacetimes, the principal null direction is geodesic and shear-free; via the Mariot-Robinson theorem \cite{Stephani:2003tm,Penrose_Rindler_1984} this defines a null test Maxwell field in the spacetime, which may have some connection to the more general type N spacetimes considered in \cite{Ortaggio_2024b,Godazgar:2020zbv}. Nevertheless, a large portion of single copies (especially stationary cases) can be understood as arising from a Killing vector.

Note this not only explains the Kerr-Schild electromagnetic potential $A_\mu = \psi k_\mu (k_\nu k^\nu) = h_{\mu \nu}\xi^\nu$ satisfying Maxwell's equations on the metric $g_{\mu \nu}$, but also explains ``double Kerr-Schild" metrics utilized in \cite{Luna:2015paa,Luna:2018dpt} of the form:
\begin{equation}
    g_{\mu \nu} = \eta_{\mu \nu} + \psi k_\mu k_\nu + \phi \ell_\mu \ell_\nu, \label{doubleKS}
\end{equation}
where the usual ``double potential" construction $A_\mu = \psi k_\mu + \phi \ell_\mu$ can be understood through considering $h_{\mu \nu} \xi^\nu$.

\subsection{Maxwell equations on a base metric}

We have thus far re-iterated the well known result that any Killing vector interpreted as an electromagnetic potential satisfies Maxwell's equations with source prescription $2\pi J^\mu = \tensor{R}{^\mu_\nu}\xi^\nu$. We furthermore stated the condition for $A_\mu \equiv h_{\mu \nu}\xi^\nu$ to be gauge equivalent to $\xi_\mu$, implying that $A_\mu$ satisfies Maxwell's equations on the full spacetime $g_{\mu \nu}$. However a key property of the Kerr-Schild double copy is that the potential $A_\mu$ also satisfies Maxwell's equations on the base metric, which we seek to generalize. Consider a metric splitting of the form:
\begin{equation}
    g_{\mu \nu} = \hat{g}_{\mu \nu} + H h_{\mu \nu} \label{genpert}
\end{equation}
where $g_{\mu \nu}$ and $\hat{g}_{\mu \nu}$ satisfy Einstein's equations, and hence we call $H h_{\mu \nu}$ an ``exact perturbation" with $H$ a constant. It is readily apparent that in the limit $H \to 0$, $g_{\mu \nu} \to \hat{g}_{\mu \nu}$. Let us assume that a Killing vector $\xi^\mu$ exists on $g_{\mu \nu}$ which is an exact form with respect to $\hat{g}_{\mu \nu}$, implying $A_\mu \equiv H h_{\mu \nu}\xi^\nu$ is a Maxwell potential on $g_{\mu \nu}$. Note that $F_{\mu \nu}$ is linear in $H$, and so the following limit must be satisfied to have $A_\mu$ be a solution on $g_{\mu \nu}$ and $\hat{g}_{\mu \nu}$:
\begin{equation}
    \lim_{H\to 0}\frac{1}{H}\nabla_\beta F_{\alpha \nu}g^{\alpha \mu}g^{\beta \nu} = \frac{1}{H}\hat{\nabla}_\alpha F_{\mu \nu}\hat{g}^{\alpha \mu}\hat{g}^{\beta \nu} = \frac{1}{H}\nabla_\beta F_{\alpha \nu}g^{\alpha \mu}g^{\beta \nu}. \label{flatlimit}
\end{equation}
where $\hat{\nabla}_\mu$ is the covariant derivative on $\hat{g}_{\mu \nu}$. This limiting procedure has a strong implication on the electromagnetic source, as this implies:
\begin{equation}
    \lim_{H \to 0} \frac{1}{H} \tensor{R}{^\mu_\nu}\xi^\nu = \frac{1}{H} \tensor{R}{^\mu_\nu}\xi^\nu ,
\end{equation}
which is to say the mixed index Ricci tensor (contracted with $\xi^\mu$) must be linear in the constant $H$ for \eqref{flatlimit} to be generically satisfied. This was already an important property in the construction of Kerr-Schild double copies through \eqref{linearKS}, however our interpretation of its role is somewhat different. Rather than extracting Maxwell's equations from the mixed index Ricci tensor, the linearity of the mixed index Ricci tensor in $H$ is a necessary condition for $A_\mu$ to satisfy sourced Maxwell's equations on both $g_{\mu \nu}$ and $\hat{g}_{\mu \nu}$.

Note that even modest deformations to the form of Kerr-Schild spacetimes such as double-Kerr-Schild spacetimes \cite{Luna:2015paa} or extended-Kerr-Schild spacetimes \cite{Ett_2010} generically exhibit non-linear contributions to their mixed index Ricci tensors. Specific manifestations of double-Kerr-Schild spacetimes have however been noted to be linear in the exact perturbation \cite{Luna:2015paa}, explaining why a classical double copy is applicable. This condition is not sufficient however, as the linearity of the mixed index Ricci tensor does not play into \eqref{flatlimit} being equivalent on both spacetimes.

\section{Tidal tensors in gravity and electromagnetism}\label{sec3}

Having now demonstrated the role Killing vectors play in the Kerr-Schild double copy, we shift into an exposition on the gravito-electromagnetism (GEM) program \cite{Costa2014,Costa_2008}, another prominent analogy between general relativity and electromagnetism.

\subsection{Tidal tensor derivations}

Consider a particle with 4-velocity $u^\mu$ undergoing a Lorentz force in a curved spacetime:
\begin{equation}
    u^\nu \nabla_\nu u^\mu = \gamma \tensor{F}{^\mu_\nu}u^\nu, \label{LorentzForce}
\end{equation}
where $\gamma = q/m$ is the charge-to-mass ratio, and $F_{\mu \nu}$ is an electromagnetic field strength. Generically this encodes both the effects of gravity and an acceleration from electromagnetism, with the former effect generically vanishing when the Riemann tensor vanishes yielding a flat space Lorentz force, and the latter when the electromagnetic field strength vanishes yielding an affinely parameterized geodesic equation.

To understand the effects these forces have on two nearby identical and intially parallel particles (therefore with identical charge-to-mass ratios), we consider the deviation vector $\delta x^\mu$ spanning the two particles' paths. Extending the derivation of the geodesic deviation equation from \cite{Wald:1984rg}, we begin by considering a family of timelike particle paths under a Lorentz force, and in the region they do not intersect we construct coordinates $t,s$ via $u^\mu \partial_\mu = \partial_t$, and $\delta x^\mu \partial_\mu = \partial_s$ implying they commute (in the sense of the Lie bracket) $[u, \delta x]^\mu = 0$. We interpret $v^\mu \equiv u^\nu \nabla_\nu \delta x^\mu$ as the relative velocity of the two particles as they travel along $u^\mu$, and importantly this quantity is assumed to be zero at some point of interest $t_0$. This implies that at $t_0$ the particle paths are parallel, and the relative accelerations will encode the true departure from the parallel geodesics of Minkowski space. The relative acceleration is represented as $a^\mu \equiv u^\nu \nabla_\nu v^\mu$, and can be expanded into the form:
\begin{equation}
    \begin{aligned}
        a^\mu &= u^\alpha \nabla_\alpha (u^\nu  \nabla_\nu \delta x^\mu)\\
        &= \delta x^\alpha \nabla_\alpha (u^\nu \nabla_\nu u^\mu) + \tensor{R}{^\mu_{\beta \alpha \nu}}u^\beta u^\nu \delta x^\alpha\\
        &= \gamma \tensor{F}{^\mu_\nu}u^\alpha \nabla_\alpha \delta x^\nu + u^\nu \nabla_\alpha (\tensor{F}{^\mu_\nu}) \delta x^\alpha + \tensor{R}{^\mu_{\beta \alpha \nu}}u^\beta u^\nu \delta x^\alpha\\
    \end{aligned}
\end{equation}
where at multiple points $\mathcal{L}_u \delta x^\mu = u^\nu \nabla_\nu \delta x^\mu - \delta x^\nu \nabla_\nu u^\mu = 0$ was utilized, along with the definition of the Riemann tensor and the Lorentz force \eqref{LorentzForce}. Importantly the first term in the final equation, $\gamma \tensor{F}{^\mu_\nu}v^\nu$, which represents the Lorentz force due to the relative velocity, is zero at $t_0$ when the particles are initially parallel. Therefore, their relative accelerations are encoded by the tensors:
\begin{equation}\label{etidal}
    \tensor{E}{^\mu_\alpha} \equiv u^\nu \nabla_\alpha \tensor{F}{^\mu_\nu}, \quad  \tensor{\mathbb{E}}{^\mu_\alpha} \equiv \tensor{R}{^\mu_{\beta \alpha \nu}}u^\beta u^\nu, 
\end{equation}
known as the electric tidal tensor and gravito-electric tidal tensor respectively. Note however that the effects of gravity and electromagnetism cannot generally be cleanly separated. Unless it is a ``test" (non-sourcing) electromagnetic field, the electromagnetic field will affect curvature via Einstein's equations. Conversely the curvature of spacetime enters the electric tidal tensor $\tensor{E}{^\mu_\alpha}$ through the covariant derivative $\nabla_\mu$. This naturally makes sense, the energy content of the EM field affect curvature, while the curvature itself dictates paths of particles.

The other set of tidal tensors can be understood as affecting extended bodies with some intrinsic spin, like a gyroscope in the case of gravity or a magnetic dipole in the case of electromagnetism. Below we will review the broad strokes of the construction, and refer readers to \cite{Dixon1973-xj,Mathisson2010-tf,Costa_2016} for further details. Consider a test body under the influence of a background (sourcing) electromagnetic field with 4-velocity $u^\mu$. Take it to only have magnetic dipole, and expand the stress energy tensor up to dipole order, integrated over the hypersurface orthogonal to $u^\mu$. One then arrives at its momentum $\mathtt{p}^\mu$, angular momentum $S^{\mu \nu}$, and magnetic dipole 2-form $\mu_{\alpha \beta}$ \cite{Mathisson2010-tf}, which satisfy the equation of motion\cite{Dixon1973-xj}:
\begin{equation}
    u^\alpha \nabla_\alpha \mathtt{p}^\mu = q \tensor{F}{^\mu_\nu}u^\nu + \frac{1}{2}\mu_{\alpha \beta}\nabla^\mu F^{\alpha \beta} - \frac{1}{2}\tensor{R}{^\mu_{\nu \alpha \beta}}u^\nu S^{\alpha \beta}. \label{extEOM}
\end{equation}
Since we are considering the observer to be the particle itself $u^\mu$, we choose $S_{\mu \nu}u^\nu = 0$ (the Mathisson-Pirani condition), and so we define the spin and magnetic dipole vectors as $S^\alpha$ and $\mu^\alpha$ respectively, satisfying:
\begin{equation}
    S_{\alpha \beta \rho \sigma} = \epsilon_{\alpha \beta \rho \sigma}S^\rho u^\sigma, \quad \mu_{\alpha \beta} = \epsilon_{\alpha \beta \rho \sigma}\mu^\rho u^\sigma
\end{equation}
which upon substitution into \eqref{extEOM} implies:
\begin{equation}
     u^\alpha \nabla_\alpha \mathtt{p}^\mu = q \tensor{F}{^\mu_\nu}u^\nu + \mu_\alpha B^{\alpha \mu} - S_\alpha \mathbb{B}^{\alpha \mu}
\end{equation}
The first term corresponds to the electric monopole (Lorentz force) and the second two terms are for the dipole moments defined by:
\begin{equation}\label{mtidal}
    \tensor{B}{^\mu_\alpha} \equiv u^\nu \nabla_\alpha (\star \tensor{F}{^\mu_\nu}), \quad  \tensor{\mathbb{B}}{^\mu_\alpha} \equiv \star \tensor{R}{^\mu_{\beta \alpha \nu}}u^\beta u^\nu,
\end{equation}
which we call the magnetic tidal tensor and gravito-magnetic tensor respectively, with $\star$ denoting the Hodge dual:
\begin{equation}
    \star F_{\mu \nu} \equiv \frac{1}{2}\epsilon_{\mu \nu \alpha \beta}F^{\alpha \beta}, \quad \star R_{\mu \beta \alpha \nu} \equiv \frac{1}{2}\tensor{\epsilon}{_{\mu \beta}^{\sigma \rho}}R_{\sigma \rho \alpha \nu}, \label{hodgeduals}
\end{equation}
where $\epsilon_{0123} = \sqrt{|g|}$ is the totally antisymmetric Levi-Civita tensor.

\subsection{Tidal force comparisons}\label{comparison}

To understand the structural differences between the electromagnetic and gravitational tidal tensors, we can re-express their equations of motion in terms of tidal tensors as was done in \cite{Costa2014}, although here we derive these equations in an alternative manner. We begin with Maxwell's equations in two equivalent forms:
\begin{equation}
    \begin{aligned}
        \nabla_\nu F^{\mu \nu} = 4\pi J^\mu \quad &\Leftrightarrow \quad \nabla_{[\mu} \star F_{\nu \alpha]} = \frac{2 \pi}{3} \epsilon_{\mu \nu \alpha \beta}J^\beta, \\
        \nabla_\nu (\star F^{\mu \nu}) = 0 \quad &\Leftrightarrow \quad \nabla_{[\mu} F_{\nu \alpha]} = 0,
    \end{aligned}
\end{equation}
where the right-hand set of equations can be understood as the component form of Maxwell's equations in terms of differential forms. Contracting these equations with some observer's 4-velocity $u^\mu$ allow us to re-express these equations in terms of tidal tensors:
\begin{equation}
\begin{aligned}
   u^\nu \nabla_\mu \tensor{F}{^\mu_\nu} = \tensor{E}{^\mu_\mu} &= -4 \pi J^\mu u_\mu,\\
   u^\nu \nabla_\mu (\star \tensor{F}{^\mu_\nu}) = \tensor{B}{^\mu_\mu} &= 0,\\
   \frac{1}{2} \left( u^\alpha \nabla_\nu F_{\mu \alpha} - u^\alpha \nabla_\mu F_{\nu \alpha} \right) =  E_{[\mu \nu]} &= \frac{1}{2}u^\alpha \nabla_\alpha F_{\mu \nu},\\
    \frac{1}{2} \left( u^\alpha \nabla_\nu (\star F_{\mu \alpha}) - u^\alpha \nabla_\mu (\star F_{\nu \alpha}) \right) = B_{[\mu \nu]} &= \frac{1}{2}u^\alpha \nabla_\alpha (\star F_{\mu \nu}) -2\pi \epsilon_{\mu \nu \alpha \beta } J^\alpha u^\beta.
\end{aligned}
\end{equation}
Similarly, one may begin with Einstein's equations with the Ricci scalar re-expressed in terms of the stress-energy tensor along with the first Bianchi identity:
\begin{equation}
    R_{\mu \nu} = 8\pi \left(T_{\mu \nu} - \frac{1}{2}g_{\mu \nu}\tensor{T}{^\alpha_\alpha} \right), \quad R_{[\mu \nu \alpha] \beta} = 0.
\end{equation}
With these two equations one can readily check that, for a timelike observer normalized such that $u^\mu u_\mu = -1$:
\begin{equation}
\begin{aligned}
    \tensor{R}{^\mu_{\beta \mu \nu}}u^\beta u^\nu &=  \tensor{\mathbb{E}}{^\mu_\mu} = 8\pi(T_{\mu \nu}u^\mu u^\nu + \frac{1}{2}\tensor{T}{^\alpha_\alpha}),\\
    R_{[\mu | \nu |\alpha] \beta}u^\nu u^\beta &= \mathbb{E}_{[\mu \alpha]} = 0.
\end{aligned}
\end{equation}
Meanwhile, the Bianchi identity can also have its Hodge dual be taken to yield: 
\begin{equation}
\begin{aligned}
    \star\tensor{R}{^\mu_{\beta \mu \nu}}u^\beta u^\nu = \tensor{\mathbb{B}}{^\mu_\mu} &= 0.
\end{aligned}
\end{equation}
The final analogue to the electromagnetic tidal tensors, the antisymmetric part of the gravito-magnetic tensor:
\begin{equation}
    \mathbb{B}_{[\mu \alpha]} = - \star R_{\nu [\mu \alpha] \beta}u^\nu u^\beta, \label{antiBdd}
\end{equation}
takes some more effort. Beginning with the irreducible composition of the Weyl tensor \cite{Stephani:2003tm}:
\begin{equation}
    R_{\mu \nu \alpha \beta} = W_{\mu \nu \alpha \beta} + E_{\mu \nu \alpha \beta} + G_{\mu \nu \alpha \beta},
\end{equation}
where $W_{\mu \nu \alpha \beta}$ is the Weyl tensor and the following definitions have been used:
\begin{equation}
    \begin{aligned}
E_{\mu \nu \alpha \beta} & \equiv \frac{1}{2}\left(g_{\mu \alpha} P_{\nu \beta}+g_{\nu \beta} P_{\mu \alpha}-g_{\mu \beta} P_{\nu \alpha}-g_{\nu \alpha} P_{\mu \beta}\right), \\
G_{\mu \nu \alpha \beta} & \equiv \frac{1}{12} R\left(g_{\mu \alpha} g_{\nu \beta}-g_{\mu \beta} g_{\nu \alpha}\right),\\
P_{\mu \nu} & \equiv R_{\mu \nu}-\frac{1}{4} R g_{\mu \nu}.
\end{aligned}
\end{equation}
Importantly, the left and right Hodge dual (where the right Hodge dual is applied to the opposite indices in \eqref{hodgeduals}) are equivalent for $W_{\mu \nu \alpha \beta}$ and $G_{\mu \nu \alpha \beta}$, meaning their antisymmetric portions in \eqref{antiBdd} are zero. Meanwhile the left and right Hodge duals of $P_{\mu \nu \alpha \beta}$ are negative of one another implying:
\begin{equation}
    \mathbb{B}_{[\mu \alpha]} = \star E_{\mu \nu \alpha \beta}u^\nu u^\beta = \frac{1}{2} \epsilon_{\mu \nu \alpha \sigma}\tensor{P}{^\sigma_\beta}u^\beta u^\nu = 4\pi \epsilon_{\mu \alpha \sigma \nu}\tensor{T}{^\sigma_\beta}u^\beta u^\nu
\end{equation}

\subsection{Conditions for exact analogies}

Having established that the tidal tensors within gravity and electromagnetism generically have different behavior, we note however, that there are concrete examples where an electromagnetic potential derived from a spacetime metric results in equivalent tidal tensors between the two theories. In this section we lay out what conditions must be generally satisfied for such a correspondence to occur.

The first immediate condition comes from the gravito-electric tidal tensor being totally symmetric; for the electric tidal tensor to also be totally symmetric it requires:
\begin{equation}
    E_{[\mu \alpha]} = \mathbb{E}_{[\mu \alpha]} \quad \implies \quad u^\alpha \nabla_\alpha F_{\mu \nu} = 0,
\end{equation}
meaning that $F_{\mu \nu}$ does not change as it is propagated along $u^\alpha$. Utilizing that $\nabla_\sigma \epsilon_{\mu \nu \alpha \beta} = 0$ (with metric compatible $\nabla_\sigma$), this furthermore implies:
\begin{equation}
    u^\alpha \nabla_\alpha (\star F_{\mu \nu}) = \star (u^\alpha \nabla_\alpha F_{\mu \nu}) = 0.
\end{equation}
This is quite a constraining condition, as it implies that the electromagnetic field must be unchanging along the observer's wordline. Conveniently however, this just leaves relations to be placed between the electromagnetic and gravitational sources. Beginning with equating the trace of the electric and the gravito-electric tidal tensors: 
\begin{equation}
    \tensor{E}{^\mu_\mu} = \tensor{\mathbb{E}}{^\mu_\mu} \quad \implies \quad J^\mu u_\mu = -\frac{1}{4\pi} R_{\mu \nu}u^\mu u^\nu = -2T_{\mu \nu}u^\mu u^\nu - \tensor{T}{^\alpha_\alpha}. \label{rhoc}
\end{equation}
To obtain the full source correspondence, and not just the charge density for observer $u^\mu$, we equate the antisymmetric portions of the magnetic tidal tensor and the gyroscopic tensor. Using the following identity:
\begin{equation}
    \epsilon_{\alpha \beta \delta \gamma}\epsilon^{\delta \gamma \mu \nu} \sim \delta_\alpha^{[\mu}\delta_\beta^{\nu]},
\end{equation}
to remove the epsilon tensors present, we set the Hodge dual of $B_{\alpha \beta}$ and $\mathbb{B}_{\alpha \beta}$ equal resulting in:
\begin{equation}
    B_{[\mu \alpha]} = \mathbb{B}_{[\mu \alpha]} \quad \implies \quad J^{[\mu}u^{\nu]} = -2\tensor{T}{^{[\mu}_\sigma}u^{\nu]}u^\sigma.
\end{equation}
Contracting with $u_\nu$ and using $u^\mu u_\mu = -1$ we get:
\begin{equation}
    -J^\mu - J^\nu u_\nu u^\mu = 2\tensor{T}{^\mu_\sigma}u^\sigma + 2 T_{\nu \sigma}u^\nu u^\sigma u^\mu,
\end{equation}
and using \eqref{rhoc}, we obtain:
\begin{equation}
    J^\mu = -2(\tensor{T}{^\mu_\sigma}u^\sigma - \frac{1}{2}\tensor{T}{^\alpha_\alpha}u^\mu) = -\frac{1}{4\pi}\tensor{R}{^\mu_\nu}u^\nu. \label{Jform}
\end{equation}
This is very reminiscent of the source correspondence present in the Kerr-Schild double copy \cite{Monteiro:2014cda}, where sourced prescriptions rely on the proposed electromagnetic field having a source proportional to a Killing vector contracted with the Ricci tensor. It is also worth noting that this same source correspondence \eqref{Jform} has been instrumental for generalizing classical double copies, for instance in generalized curved backgrounds \cite{Carrillo-Gonzalez:2017iyj} and in ascertaining the form of a sourced Weyl double copy \cite{Easson:2021asd,Easson:2022zoh}, albeit with $u^\mu = \xi^\mu$ being a Killing vector.

\section{Killing vectors and tidal analogies}\label{sec4}

Any example exhibiting exact tidal analogies hence must have two prominent features: the EM field does not change along the observer's worldline, and the sources have the same structure as present in the classical double copy \cite{Monteiro:2014cda}. The former observation is reconciled by the fact that typical examples exhibiting tidal analogies \cite{Costa_2008} have been formed from stationary spacetimes with the observer themselves following the timelike Killing vector $u^\mu \partial_\mu = \partial_t$. It is therefore natural to ask how central Killing vectors are to such analogies.

In this section, we establish the properties of Killing vectors regarding tidal analogies, demonstrating that if one chooses an observer $u^\mu = \xi^\mu$ and field strength $F_{\mu \nu} \sim \nabla_\mu \xi_\nu$ for some Killing vector $\xi^\mu$, ``tidal analogies" are automatically satisfied. There is of course a large caveat in having an observer aligned with a Killing vector, in that $\xi^\mu$ must be timelike and geodesic to be a physical observer, and otherwise \eqref{etidal} are not truly tidal tensors. We show however that typical examples exhibiting tidal analogies are gauge equivalent to choosing a timelike, geodesic Killing vector $\xi_\mu$ as the electromagnetic potential, and $\xi^\mu$ as the observer.

\subsection{Killing vector potentials for Tidal tensors, Kerr-Schild double copy}\label{killingtidal}

Recall that for an electromagnetic field generated by a Killing vector, we have the relation:
\begin{equation}
    \nabla_\mu F_{\alpha \beta} = 2\nabla_\mu \nabla_\alpha \xi_\beta = 2\xi_\nu\tensor{R}{^\nu_{\mu \alpha \beta}}. \label{xirelation}
\end{equation}
This implies that the Lie derivative of $F_{\mu \nu}$ along $\xi^\mu$ is zero, since:
\begin{equation}
\begin{aligned}
    \mathcal{L}_\xi F_{\mu \nu} &= \xi^\alpha \nabla_\alpha F_{\mu \nu} + F_{\alpha \nu}\nabla_\mu \xi^\alpha + F_{\mu \alpha}\nabla_\nu \xi^\alpha \\
    &= \xi^\alpha \nabla_\alpha F_{\mu \nu} + (\nabla_\nu \xi_\alpha)(\nabla_\mu \xi^\nu) + (\nabla_\alpha \xi_\mu)(\nabla_\nu \xi^\alpha)\\
    &= \xi^\alpha \nabla_\alpha F_{\mu \nu} + (\nabla_\nu \xi_\alpha)(\nabla_\mu \xi^\alpha) - (\nabla_\mu \xi_\alpha)(\nabla_\nu \xi^\alpha)\\
    &= \xi^\alpha \nabla_\alpha F_{\mu \nu}, \label{liecalc}
\end{aligned}
\end{equation}
and from \eqref{xirelation}:
\begin{equation}
    \xi^\alpha \nabla_\alpha F_{\mu \nu} =  2\xi^\alpha \xi^\beta R_{\alpha \beta \mu \nu} = 0.
\end{equation}
Importantly, for any (conformal) Killing vector $\xi$, the Hodge dual of a 2-form ``commutes" with the Lie derivative \cite{Huggett:1986fs}, that is:
\begin{equation}
    \star \left(\mathcal{L}_\xi F_{\mu \nu} \right) = \mathcal{L}_\xi (\star F_{\mu \nu}),
\end{equation}
which implies $\mathcal{L}_\xi (\star F_{\mu \nu}) = 0$. Additionally as previously noted, just utilizing $\nabla_\sigma \epsilon_{\mu \nu \alpha \beta} = 0$ implies:
\begin{equation}
    \xi^\alpha \nabla_\alpha (\star F_{\mu \nu}) = 0.
\end{equation}
We can furthermore calculate the electric and magnetic tidal tensors. The electric tidal tensor is proportional to the electric part of the Riemann tensor, as follows immediately from \eqref{xirelation}:
\begin{equation}\label{getidal}
    \xi^\nu \nabla_\alpha F_{\mu \nu} = -2R_{\mu \beta \alpha \nu}\xi^\beta \xi^\nu.
\end{equation}
After taking the Hodge dual, we arrive at:
\begin{equation}
    \nabla_\alpha(\star F_{\mu \nu}) = -2 \star R_{\mu \nu \alpha \beta}\xi^\beta.
\end{equation}
It follows that the magnetic part of the Riemann tensor, and magnetic tidal tensors are proportional\footnote{One can of course re-scale the field strength $F_{\mu \nu}$ if one seeks to remove the $-2$.}:
\begin{equation}\label{gmtidal}
    \xi^\nu \nabla_\alpha(\star F_{\mu \nu}) = -2 \star R_{\mu \beta \alpha \nu}\xi^\beta\xi^\nu.
\end{equation}
In short the properties of Killing vectors ensure that along the Killing vector, the gravitational ``tidal tensors" are equivalent to ``tidal tensors" formed from the two-form defined from said Killing vector. However for these structures to truly represent tidal tensors, we must have $\xi^\mu$ be timelike and geodesic.

\subsection{Examples of exact tidal analogies}
In section \ref{comparison} we noted how the electromagnetic and gravitational tidal tensors are generally very different but exhibit the same symmetries when the electromagnetic fields are stationary in the rest frame of the observer. However, in two special cases observed in \cite{Costa_2008}, this analogy becomes exact, i.e. the tidal tensors on the two sides match by virtue of identifying some components of the background metric with an electromagnetic potential. In this section, we explicitly show that these potentials are gauge equivalent to the timelike Killing vector arising naturally in these geometries, thereby rendering known examples of exact tidal analogies as a consequence of this gauge equivalence using \eqref{getidal},\eqref{gmtidal}. 

\subsubsection*{Ultra-stationary spacetimes}

Ultra-stationary spacetimes are stationary spacetimes that may be chosen to have $g_{00}$ constant in the chart where the metric is explicity time independent, with the general line element corresponding to:
\begin{equation}
    ds^2 = -(dt+A_i(x_k) dx^i)^2 + (\delta_{ij}+2h_{ij}(x_k))dx^i dx^j. \label{us}
\end{equation}
The G\"odel spacetime \cite{Godel:1949ga}, the Som-Raychaudhuri metrics \cite{Som} and the metric of Lense and Thirring \cite{lense-thirring} are all examples of ultra-stationary spacetimes.

It was observed in \cite{Costa_2008} that when $(0,\vec{A})$ is identified  with an electromagnetic potential $A_\mu$ living on the same background, the gravitational tidal tensors agree exactly with the electromagnetic ones.   
That the field strength of the timelike Killing vector $\xi^\mu = (1,0,0,0)$ equals that of the electromagnetic potential up to a minus sign, follows almost immediately upon writing $\xi_\mu=(-1,-\vec{A})$.
Then, the field strength tensor coming from $-\xi_\mu$ which we denote by $F^\xi_{\mu\nu}$ becomes
\begin{equation}
        F^\xi_{\mu\nu} \equiv \nabla_\mu(-\xi_\nu)-\nabla_\nu(-\xi_\mu) = \nabla_\mu(A_\nu)-\nabla_\nu(A_\mu)\equiv F^A_{\mu\nu}.
\end{equation}
This then further reduces to the field strength on the spatial three metric, $F_{ij}\equiv\widetilde{\nabla}_i A_j - \widetilde{\nabla}_j A_i$ since $\vec{A}$ is independent of time, where $\widetilde{\nabla}_k$ are covariant derivatives defined on $g_{ij}=\delta_{ij}+2h_{ij}(x_k)$.
This feature of ultra-stationary spacetimes is not true in general, for instance general stationary spacetimes given by: 
\begin{equation}\label{stat}
    ds^2 = -e^{\phi(x_k)}(dt+A_i(x_k) dx^i)^2 + (\delta_{ij}+2h_{ij}(x_k)) dx^i dx^j. 
\end{equation}
The field strength coming from the Killing vector of stationary spacetimes can no longer be reduced to one on the spatial metric since $\xi_\mu$ now gains a non-trivial timelike component due to $\phi(x_k)$.

Furthermore, that $\xi^\mu$ is geodesic and normalizable such that $\xi^\mu \xi_\mu = -1$ is important for the timelike Killing vector to represent an observer. Hence the equations \eqref{getidal}, \eqref{gmtidal} arising from a Killing vector are indeed interpretable as tidal tensors.

\subsubsection*{Linearized gravito-electromagnetic analogies}

The metric describing arbitrary, linear order perturbations around Minkowski can be given by:
\begin{equation}\label{lin}
    ds^2= -\left(1+\phi(t,x^i)\right)dt^2 -2A_j(t,x^i)dt dx^j+(\delta_{ij} +2h_{ij}(t,x^i))dx^i dx^j.
\end{equation}
We will be interested in the metric for stationary linearized gravity, i.e. when $\phi, A_j$ and $h_{ij}$ are time independent, which can be obtained from the metric for stationary spacetimes \eqref{stat} by expanding to linear order in $\phi$ and $\vec{A}$. It was observed in \cite{Costa_2008} that for time independent perturbations, and a static observer, i.e. with four velocity $u^\mu=\tensor{\delta}{^\mu_0}$, the gravitational tidal tensors agree with the electromagnetic ones upon identifying $(\phi,A_i)$ with an electromagnetic potential $A_{\mu}$, with the tidal tensors \eqref{etidal} and \eqref{mtidal} taking the form \cite{Costa_2009}:
\begin{equation}
    \mathbb{E}_{ij} \simeq \frac{1}{2}\partial_{i}\partial_j\phi=E_{ij},\;\mathbb{B}_{ij}\simeq \frac{1}{2}\tensor{\epsilon}{_i^{lk}}\partial_l\partial_j A_k =B_{ij}.
\end{equation} 
The electromagnetic tidal tensors computed on the curved background reduce to the ones on the flat background since working up to linear order in the perturbations reduces covariant derivatives to partial derivatives on the three dimensional flat metric. This means that the electromagnetic fields obtained from the potential $A_\mu$, also live on the flat metric. 
In this sense, working in the limit of linearized gravity perturbations effectively decouples gravity from the tidal effects of electromagnetism.  

For time independent perturbations, $\xi^\mu=(1,0,0,0)$ is the stationary Killing vector of \eqref{lin}. The field strength corresponding to $-\xi_\mu=-(-1- \phi,-A_i)$ given by $F^\xi_{\mu\nu} \equiv -2\nabla_\mu\xi_\nu$ becomes 
\begin{equation}
    F^\xi_{\mu\nu} = F^A_{\mu\nu} = \partial_\mu A_\nu-\partial_\nu A_\mu.
\end{equation}
Thus, the electromagnetic potential identified under the analogy is once again gauge equivalent to the timelike Killing vector. 

\subsection{Connections with double copy, physical interpretations}

We have shown that the equivalence of electromagnetic and gravitational tidal tensors in these cases is therefore a consequence of Killing's equation. Importantly, the Killing vector in these example is a timelike geodesic, and so can be interpreted as being aligned with the worldline of some observer. Given that we've shown that many potentials within the Kerr-Schild double copy are gauge equivalent to Killing vectors, this implies classical double copies of this form exhibit the mathematical structure of ``exact tidal analogies" in the sense of \eqref{getidal}, although these Killing vectors must be timelike geodesics to physically represent tidal tensors. The examples within these programs are similar in that they ``generate" an electromagnetic potential from the spacetime metric which mimics the effects of electromagnetism, relying on the one-form associated with a Killing vector.

An example relating these two programs was the G\"odel double copy analyzed in \cite{Kent2025}. Although not having a Kerr-Schild metric form, a single copy was proposed for the G\"odel metric utilizing the structure of the Weyl double copy \cite{Luna:2018dpt}. The observations within this work have explained several of the correspondences in \cite{Kent2025}, for example the single copy having a source like \eqref{KillingSource} is due to the proposed electromagnetic field also being generated from the stationary Killing vector present in the spacetime. That this Killing vector is also a geodesic allowed for a direct correspondence with gravito-electromagnetism, meaning the electromagnetic analogies described in \cite{Costa_2008,Costa2014} were also applicable.

The last prominent feature of these analogies that must be discussed is the generic inability to decouple gravity within the electromagnetic tidal tensors, beyond just drawing the potentials from the metric. A prominent motivation for both the GEM and CDC programs is to have an electromagnetic field on flat space mimic some structure of relativity on curved space. Kerr-Schild metrics serve as a special class of spacetimes where the electromagnetic field strength may be framed as satisfying Maxwell's equations on the flat background metric, however that is not true for the tidal tensors built from said field strengths. Therefore the acceleration of nearby particles cannot be described merely using a tidal tensor on a flat background, which perhaps should not be surprising. One can of course force this to be true by only considering linear order perturbations, in which case the covariant derivatives action on anything containing the electromagnetic potential (which itself is drawn from the metric) will reduce to just flat space derivatives.

\section{Connecting the Kerr-Schild and Weyl double copies}\label{sec5}

Another prominent formulation of the classical double copy employs the spinor formalism of general relativity to provide a coordinate-free approach to understanding the structure of double copies. Specifically, the program is based upon observations that Type D vacuum spacetimes can have their Weyl spinor decomposed into a symmetrized product of electromagnetic spinors \cite{Walker:1970un,Hughston:1972qf}, which provides a spinorial approach to the classical double copy \cite{Luna:2018dpt}. This procedure has been extended to sourced spacetimes \cite{Easson:2021asd,Easson:2022zoh}, which exhibit a signature splitting of the Weyl spinor into a sum over symmetrized products of separate sourced electromagnetic spinors, which has been derived at linearized level utilizing twistors \cite{Armstrong-Williams:2024bog}. At the level of exact solutions this has remained unexplained, and so too has a general understanding of the connection between the Weyl and Kerr-Schild double copies.

Killing vectors must be central in this connection since the Kerr-Schild double copy, which is equivalent to the prominent examples within the Weyl double copy, is reliant upon them. One such class of spacetimes, those containing a Killing-Yano tensor, serve as a prominent example demonstrating the relationship between field strengths built from a unique Killing vector and the Weyl tensor. Although this class of spacetimes contains many prominent Weyl double copy examples, they are not meant to be exhaustive, meaning there are spacetimes analyzed within the Weyl double copy outside of this class. Rather, they serve as a proof of concept how one can connect the Kerr-Schild and Weyl double copies. We provide some description at the end of this section on how one might replicate such constructions for other Weyl double copies outside of this class.

\subsection{The Weyl double copy}

To further set conventions we utilize those of \cite{Stephani:2003tm} regarding spinors, since it preserves the spinor algebra defined in \cite{Penrose_Rindler_1984} while using $(-+++)$ signature. This choice of signature induces an asymmetry between passing from spacetime $\to$ spinors as opposed to spinors $\to$ spacetime. That is to say:
    \begin{equation}
        v^a = - \sigma^a_{\lh{A}\rh{A}}v^{\lh{A}\rh{A}} \ \ \longleftrightarrow \ \ v_{\lh{A}\rh{A}} = \sigma^a_{\lh{A}\rh{A}}v_a.
    \end{equation}
The canonical resource for spinors in general relativity is \cite{Penrose_Rindler_1984}, and a more succinct introduction in the context of classical double copies can be found in \cite{Kent2025}. Here we will present just enough to understand the structure of the Weyl tensor relevant for understanding the Weyl double copy.

The foundation of the spinorial formalism for general relativity uses the relation that the group of proper orthochronous Lorentz transformations $SO^+(1,3)$ is double covered by $SL(2,\mathbb{C})$. Given a 4-vector on Minkowski space $v^a$, we can instead represent it via a change of basis as:
\begin{equation}
    v^a \sigma_a^{\lh{A}\rh{A}} = \begin{pmatrix}
        v^0 + v^3&v^1 + iv^2\\
        v^1-iv^2&v^0-v^3
    \end{pmatrix} \equiv v^{\lh{A}\rh{A}}.
\end{equation}
The spinor indices are related by complex conjugation (that is, ``left-handed" indices $A$ are turned into ``right-handed" indices $\rh{A}$ upon complex conjugation). They are furthermore raised and lowered with respect to the $SL(2,\mathbb{C})$ invariant, totally antisymmetric Levi-Civita symbol:
\begin{equation}
    \epsilon_{\lh{A}\lh{B}} = \begin{pmatrix}
        0&1\\
        -1&0
    \end{pmatrix} = \epsilon^{\lh{A}\lh{B}},
\end{equation}
with raising and lowering conventions ``raise on the right" and ``lower on the left" for $\epsilon$:
\begin{equation}
    \kappa^{\lh{A}} = \epsilon^{\lh{A}\lh{B}}\kappa_B, \quad \kappa_{\lh{B}} = \kappa^{\lh{A}}\epsilon_{\lh{A}\lh{B}}.
\end{equation}
And so we see that the inner product is represented as:
\begin{equation}
    \eta_{ab}v^av^b = -\epsilon_{\lh{A}\lh{B}}\epsilon_{\rh{A}\rh{B}}v^{\lh{A}\rh{A}}v^{\lh{B}\rh{B}} = -\det(v^{\lh{A}\rh{A}}),
\end{equation}
with the manifestation of the double cover being the invariance under $v^{\lh{A}\rh{A}}\to-v^{\lh{A}\rh{A}}$. The fact that at any given point, the metric on the tangent space in GR is isomorphic to Minkowski space implies that at every point we can perform such a transformation, generically called the \emph{vierbein} formalism. Without a coordinate chart, different tangent spaces are required to follow the Cartan structure equations, which are in turn related to the spin-coefficient formalism. For our purposes, we will mainly concern ourselves with the structure at a single point rather than the differential structure.

Since we can utilize vierbeins to convert a coordinate basis into a Minkowski basis, it is therefore possible to go directly from a coordinate basis to spin space via the soldering forms:
\begin{equation}
    v_\mu \sigma^{\mu}_{\lh{A}\rh{A}} = v_{\lh{A}\rh{A}},
\end{equation}
where we write all spacetime indices in the coordinate basis with Greek indices. Two important facts make bivectors particularly simple in the spinor formalism. Consider a bivector $F_{\mu\nu}$ rewritten as:
\begin{equation}
F_{\mu\nu}\sigma^{\mu}_{\lh{A}\rh{A}}\sigma^\nu_{\lh{B}\rh{B}} = F_{\lh{A}\rh{A}\lh{B}\rh{B}} = F_{[\lh{A}\lh{B}]\rh{A}\rh{B}} + F_{\lh{A}\lh{B}[\rh{A}\rh{B}]}.
\end{equation}
All antisymmetric rank-2 spinors are proportional to $\epsilon_{\lh{A}\lh{B}}$, generally any totally antisymmetric $D$-dimensional tensor is proportional to $\epsilon_{a_1...a_D}$ in $D$-dimensions, with all higher valence totally antisymmetric tensors being zero. Therefore:
\begin{equation}
 F_{\mu\nu}\sigma^{\mu}_{\lh{A}\rh{A}}\sigma^\nu_{\lh{B}\rh{B}} = f_{\lh{A}\lh{B}}\epsilon_{\rh{A}\rh{B}} + \overline{f}_{\rh{A}\rh{B}}\epsilon_{\lh{A}\lh{B}}, \label{MaxwellGen}
\end{equation}
for some $f_{\lh{A}\lh{B}}$ with $f_{\lh{A}\lh{B}} = f_{(\lh{A}\lh{B})}$. The second important fact is that any totally symmetric spinor can be decomposed into a symmetrized product of rank-1 spinors:
\begin{equation}
 f_{\lh{A}\lh{B}} = \alpha_{(A}\beta_{B)},\label{fAB}
\end{equation}
where $\alpha^A$ and $\beta^B$ are called the \emph{principal null spinors} of $f_{\lh{A}\lh{B}}$. The null vectors formed from the principal null spinors are called \emph{principal null directions}, and are the eigenvectors of $F_{\mu \nu}$. Furthermore under the hodge dual, any rank-2 tensor $F_{\mu \nu}$ can be decomposed into self-dual ($\star F^+_{\mu\nu} = iF^+_{\mu\nu}$) and anti-self-dual ($\star F^-_{\mu\nu} = -iF^-_{\mu\nu}$) pieces. These pieces have the nice form:
\begin{equation}
 F^-_{\mu\nu}\sigma^{\mu}_{\lh{A}\rh{A}}\sigma^\nu_{\lh{B}\rh{B}} = f_{\lh{A}\lh{B}}\epsilon_{\rh{A}\rh{B}}, \qquad F^+_{\mu\nu}\sigma^{\mu}_{\lh{A}\rh{A}}\sigma^\nu_{\lh{B}\rh{B}} =\overline{f}_{\rh{A}\rh{B}}\epsilon_{\lh{A}\lh{B}}. \label{selfdualspin}
\end{equation}
The Weyl tensor, based upon its symmetries and being trace-free in all indices, decomposes in spinor form as:
\begin{equation}
    W_{\mu \nu \alpha \beta} \to \Psi_{\lh{A}\lh{B}\lh{C}\lh{D}}\epsilon_{\rh{A}\rh{B}}\epsilon_{\rh{C}\rh{D}} + \Bar{\Psi}_{\rh{A}\rh{B}\rh{C}\rh{D}}\epsilon_{\lh{A}\lh{B}}\epsilon_{\lh{C}\lh{D}},
\end{equation}
where $\Psi_{\lh{A}\lh{B}\lh{C}\lh{D}}$ is totally symmetric in its indices, implying:
\begin{equation}
    \Psi_{\lh{A}\lh{B}\lh{C}\lh{D}} = \alpha_{(\lh{A}}\beta_{\lh{B}}\gamma_{\lh{C}}\delta_{\lh{D})}.
\end{equation}
This gives a particularly simple formulation of the \emph{Petrov classification}, where spacetimes at each point (which can be extended globally) can be classified by how many of the Weyl tensor's principal null directions are aligned. Of particular interest for the classical double copy are \emph{Type D} spacetimes, for which two sets of two spinors are aligned, implying:
\begin{equation}
    \psi_{\lh{A}\lh{B}\lh{C}\lh{D}} = \alpha_{(\lh{A}}\beta_{\lh{B}}\alpha_{\lh{C}}\beta_{\lh{D})}.
\end{equation}
This already is in a suggestive form of ``squaring" some \eqref{fAB} to form the Weyl spinor, however the differential structure has not been included. It was proven in \cite{Walker:1970un,Hughston:1972qf} that vacuum Type D spacetimes can always have their Weyl tensors decomposed into the form:
\begin{equation}
    \psi_{\lh{A}\lh{B}\lh{C}\lh{D}} = \frac{1}{S}f_{(\lh{A}\lh{B}}f_{\lh{C}\lh{D})}, \label{WDC}
\end{equation}
where the bivector formed from $f_{\lh{A}\lh{B}}$ satisfies Maxwell's vacuum equations. These metrics always can be placed into a ``double-Kerr-Schild" form \cite{plebanski1976rotating} which has allowed a direct comparison with the Kerr-Schild double copy \cite{Luna:2018dpt}, demonstrating an equivalence in these cases and further the observation that $S$ satisfies a wave equation on the flat space of the Kerr-Schild representation. Hence analyzing the electromagnetic theories and structure from analyzing the decomposition of the Weyl spinor into electromagnetic parts has taken the name of the \emph{Weyl double copy}.

That the Kerr-Schild double copy allowed for extensions to sourced spacetimes begged the question of whether the Weyl double copy remained true, and if so how it manifested sources. The question was investigated in \cite{Easson:2021asd,Easson:2022zoh} where it was noted that the Weyl spinor split into a sum of vacuum and sourced pieces:
\begin{equation}
    \psi_{\lh{A}\lh{B}\lh{C}\lh{D}} = \frac{1}{S^{(0)}}f^{(0)}_{(\lh{A}\lh{B}}f^{(0)}_{\lh{C}\lh{D})} + \sum_i \frac{1}{S^{(i)}}f^{(i)}_{(\lh{A}\lh{B}}f^{(i)}_{\lh{C}\lh{D})} \label{sourcedWDC}
\end{equation}
with $f^{(0)}_{\lh{A}\lh{B}}$ satisfying Maxwell's vacuum equations and each $f^{(i)}_{\lh{A}\lh{B}}$ being sourced by a term in the Ricci tensor. This correspondence was made by connecting the Weyl double copy to the Kerr-Schild double copy, and since Killing vectors are important for the latter, we must understand the structure of Killing vectors at the level of the Weyl tensor for such spacetimes in order to connect them to the Weyl double copy. A large class of spacetimes that have been investigated admit \emph{Killing-Yano tensors}, so we utilize known properties of such spacetimes to explain the structure of the Weyl double copy within them. We also note that in higher dimensional spacetimes, Killing-Yano tensors have also been utilized to glean further structure \cite{Chawla:2022ogv} within the context of the Weyl double copy, further strengthening the correspondence drawn here.

\subsection{Spacetimes admitting Killing-Yano tensors}

This section relies heavily on the analysis of spacetimes containing Killing-Yano tensors performed in \cite{Dietz}. For full proofs we point readers to that work; we wish to instead utilize the known properties of such spacetimes to connect Killing vectors to the Weyl double copy.

A Killing-Yano tensor is a generalization of Killing's equation for higher valence tensors. Specifically, given a bivector $\varphi_{\mu \nu}$, a generalization of Killing's equations satisfying:
\begin{equation}
    \nabla_{(\mu}\star \varphi_{\nu)\alpha} = 0,
\end{equation}
is defined as a \emph{Killing-Yano tensor}, where the inclusion of the Hodge dual $\star$ is a convenient choice simplifying later expressions. Defining the vector:
\begin{equation}
    \xi^\alpha = \frac{1}{3} \nabla_\mu \varphi^{\mu \alpha}, \label{xidef}
\end{equation}
implies the relation:
\begin{equation}
    \nabla_\alpha (\star \varphi_{\mu \nu}) = \epsilon_{\mu \nu \alpha \beta}\xi^\beta.
\end{equation}
It is more convenient to consider the anti-self-dual bivector:
\begin{equation}
    \Phi_{\mu \nu} \equiv \varphi_{\mu \nu} + i \star \varphi_{\mu \nu},
\end{equation}
which satisfies:
\begin{equation}
    \nabla_\alpha \Phi_{\mu \nu} = 2 \xi_{[\nu}g_{\mu]\alpha} - i\epsilon_{\mu \nu \alpha \beta}\xi^\beta. \label{phieqn}
\end{equation}
How such spacetimes relate to the Riemann tensor requires applying subsequent covariant derivatives to \eqref{phieqn} and applying the definition of the Riemann tensor. We assume that $\varphi_{\mu \nu}$ is algebraically general, that is it has two unique principal null spinors when decomposed as \eqref{fAB}. In that case, $\xi^\mu$ is a Killing vector:
\begin{equation}
    \nabla_{(\mu}\xi_{\nu)} = 0,
\end{equation}
and upon relating \eqref{phieqn} to the Riemann tensor, the following integrability condition is satisfied:
\begin{equation}
    2\nabla_{\mu}\xi_\nu + i \epsilon_{\mu \nu \alpha \beta}\nabla^\alpha \xi^\beta = \frac{1}{2}W_{\mu \nu \alpha \beta}\Phi^{\alpha \beta}+\frac{1}{6}R \Phi_{\mu \nu}. \label{intcond}
\end{equation}
We may define the bivector induced from the Killing vector as $F_{\mu \nu} \equiv 2 \nabla_\mu \xi_\nu$, and furthermore note that any anti-self-dual bivector has the form:
\begin{equation}
    \Phi_{\mu \nu}\sigma^\mu_{\lh{A}\rh{A}}\sigma^\nu_{\lh{B}\rh{B}} \equiv \chi_{\lh{A}\lh{B}}\epsilon_{\rh{A}\rh{B}} \equiv \phi \, o_{(\lh{A}}\iota_{\lh{B})}\epsilon_{\rh{A}\rh{B}}, \label{Phidef}
\end{equation}
where we have chosen a spinor basis aligned with the principal null directions of $\Phi_{\mu \nu}$ such that:
\begin{equation}
    o_{\lh{A}} \iota^{\lh{A}} = -o^{\lh{A}} \iota_{\lh{A}} = 1,\ \ o_{\lh{A}} o^{\lh{A}} = \iota_{\lh{A}} \iota^{\lh{A}} = 0.
\end{equation}

Importantly, all spacetimes admitting Killing-Yano tensors that are algebraically general (two unique principal null spinors) are necessarily Type D, and furthermore these spinors $o_{\lh{A}}$ and $\iota_{\lh{A}}$ are also principal null spinors of the Weyl spinor. That is to say, the Weyl spinor decomposes as:
\begin{equation}
    \Psi_{\lh{A}\lh{B}\lh{C}\lh{D}} = 6 \, \Psi_2 \, o_{(\lh{A}}\iota_{\lh{B}}o_{\lh{C}}\iota_{\lh{D})}.
\end{equation}
From here we need only note that:
\begin{equation}
    2\nabla_{\mu}\xi_\nu + i \epsilon_{\mu \nu \alpha \beta}\nabla^\alpha \xi^\beta = F_{\mu \nu} + i\star F_{\mu \nu} \equiv F^-_{\mu \nu},
\end{equation}
which can be written as:
\begin{equation}
    F^-_{\mu \nu}\sigma^\mu_{\lh{A}\rh{A}}\sigma^\nu_{\lh{B}\rh{B}} \equiv f_{\lh{A}\lh{B}}\epsilon_{\rh{A}\rh{B}},
\end{equation}
in order to fully express \eqref{intcond} into spinor form:
\begin{equation}
    f_{\lh{A}\lh{B}}\epsilon_{\rh{A}\rh{B}} = 3 \left( \Psi_2 \, o_{(\lh{A}}\iota_{\lh{B}}o_{\lh{C}}\iota_{\lh{D})} \epsilon_{\rh{A}\rh{B}}\epsilon_{\rh{C}\rh{D}} + \Bar{\Psi}_2 \, o_{(\rh{A}}\iota_{\rh{B}}o_{\rh{C}}\iota_{\rh{D})}\epsilon_{\lh{A}\lh{B}}\epsilon_{\lh{C}\lh{D}} \right)\phi \, o^{(\lh{C}}\iota^{\lh{D})} + \frac{1}{6}R\phi \, o_{(\lh{A}}\iota_{\lh{B})}.
\end{equation}
After contracting and removing $\epsilon_{\rh{C}\rh{D}}$ from each side, we arrive at the remarkably simple form:
\begin{equation}
    f_{\lh{A}\lh{B}} = \phi (\frac{R}{6} - \Psi_2) o_{(\lh{A}}\iota_{\lh{B})}. \label{finalfAB}
\end{equation}
This explains many observations of the correspondence between the Weyl and the Kerr-Schild double copies for such spacetimes. First and foremost, since the Kerr-Schild double copy is gauge equivalent to the Killing vector $\xi_\mu$, for the Kerr-Schild double copy to have a direct relation to a Weyl double copy requires the bivector $F_{\mu \nu}$ to have some relation to the Weyl tensor, which is achieved through \eqref{intcond}. Secondly, treatments of the sourced Weyl double copy have taken the spacetime to have no cosmological constant \cite{Easson:2021asd,Easson:2022zoh} which is clear from \eqref{finalfAB}, as $R=0$ is required for a direct correspondence between the Weyl tensor and $F_{\mu \nu}$. This can be understood in the following way as well; the Weyl spinor can be rewritten from the 2nd Bianchi identity into the form:
\begin{equation}
    \nabla^{\lh{A}}_{\rh{B}}\Psi_{\lh{A}\lh{B}\lh{C}\lh{D}} = \nabla^{\rh{A}}_{(\lh{B}}P_{\lh{C}\lh{D})\rh{A}\rh{B}}, \label{spinBianchi}
\end{equation}
where $P_{\lh{A}\rh{A}\lh{B}\rh{B}}$ is the spinor analogue of the trace-free Ricci tensor, meaning that the Weyl spinor is only related to the trace-free portion of the Ricci tensor. Therefore any Weyl double copy necessarily cannot have information from the cosmological constant (or more broadly the Ricci scalar). Lastly, it is the splitting of the Weyl scalar $\Psi_2$ into multiple pieces:
\begin{equation}
    \Psi_2 = \sum_i \psi^{(i)},
\end{equation}
that directly implies a splitting of the electromagnetic spinor into multiple parts via (assuming $R=0$):
\begin{equation}
    f_{\lh{A}\lh{B}} = -\sum_i \phi \, \psi^{(i)} o_{(\lh{A}}\iota_{\lh{B})},
\end{equation}
and hence the Weyl spinor can always be rearranged into the form \eqref{sourcedWDC}.

\subsection{Examples admitting a Killing-Yano tensor}

Now that we have seen that spacetimes with a Killing-Yano tensor provide a simple correspondence between Killing vectors and the Weyl tensor, we seek to demonstrate that many prominent examples fall within this range. The Weyl double copy has included limits of the most general Type-D electrovacuum solution (the Plebanski-Demianski solution) \cite{plebanski1976rotating,Griffiths_Pleb} with aligned principal null directions between the electromagnetic field and Weyl spinor, spherically symmetric metric expansions about Schwarzschild \cite{Easson:2021asd}, or Type N gravitational waves \cite{Godazgar:2020zbv} to name a few. The Plebanski-Demianski metric is known to have both the principal null spinors of the spacetime be geodesic and shear-free \cite{Griffiths_Pleb}, which means a spinor satisfies:
\begin{equation}
    o^{\lh{A}}o^{\lh{B}}\nabla_{\lh{A}\rh{A}}o_{\lh{B}} = 0.
\end{equation}
The spherically symmetric expansion also has its principal null spinors be geodesic and shear-free, as we will show.

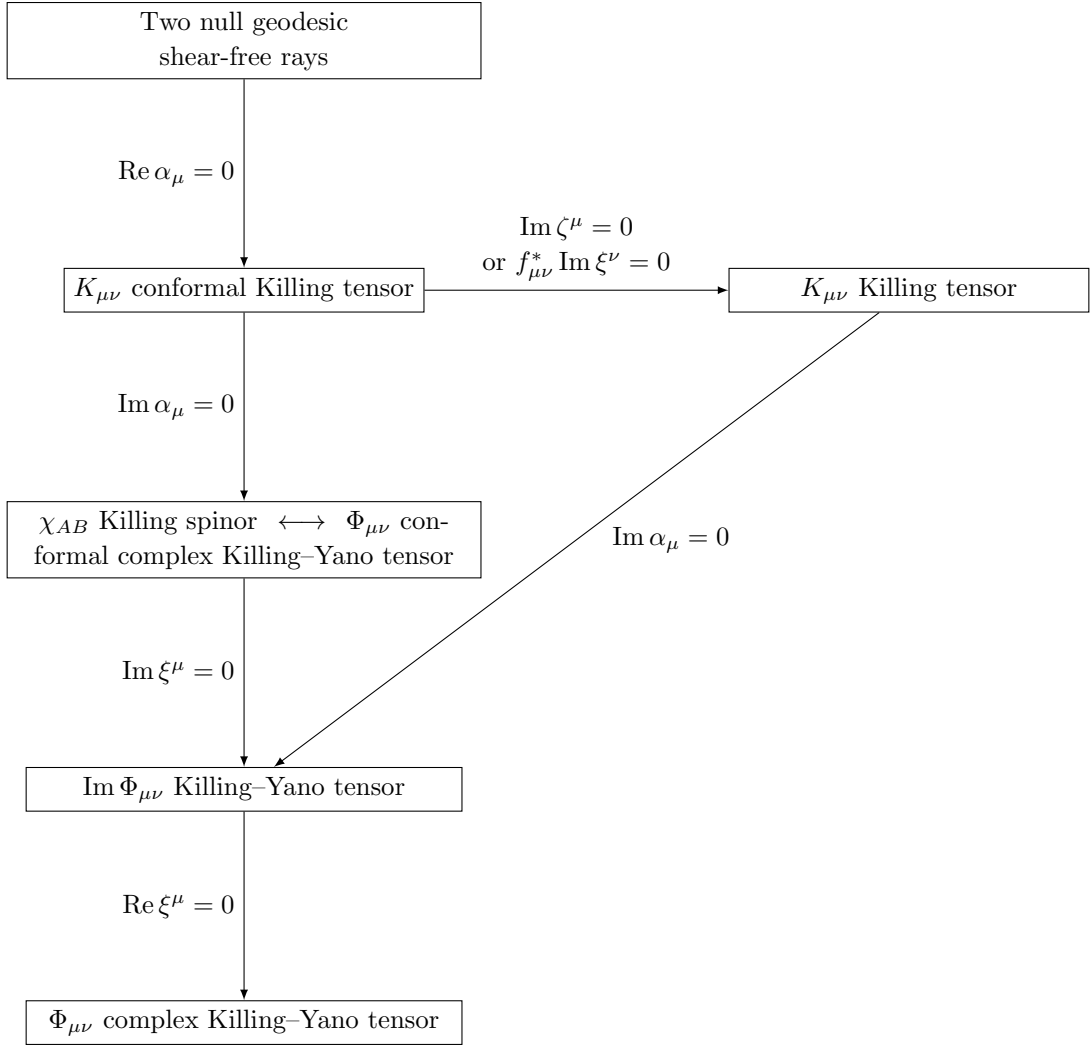
\begin{figure}
\centering

\begin{tikzpicture}[>=latex, node distance=3cm, auto]

% Top box
\node (start) [draw, rectangle, text width=6cm, align=center]
{Two null geodesic \\ shear-free rays};

% K_ab conformal Killing tensor
\node (KConf) [draw, rectangle, text width=4.5cm, align=center, below=2.5cm of start]
{$K_{\mu \nu}$ conformal Killing tensor};

% Arrow from start to KConf
\draw [->] (start) -- node[left]{\(\mathrm{Re}\,\alpha_{\mu}=0\)} (KConf);

% Box: Killing spinor / conformal complex Killing Yano tensor
\node (KillSpinor) [draw, rectangle, text width=6cm, align=center, below=2.5cm of KConf]
{$\chi_{AB}$ Killing spinor 
 $\;\longleftrightarrow\;$ 
 $\Phi_{\mu \nu}$ conformal complex Killing--Yano tensor};

% Arrow from KConf down to KillSpinor
\draw [->] (KConf) -- node[left]{\(\mathrm{Im}\,\alpha_{\mu}=0\)} (KillSpinor);

% Box: K_ab Killing tensor (branching to the right)
\node (KKill) [draw, rectangle, text width=4.5cm, align=center, right=4.0cm of KConf]
{$K_{\mu \nu}$ Killing tensor};

% Arrow from KConf to KKill
%\draw [->] (KConf) -- node[above]{\(\mathrm{Im}\,\xi^{\mu}=0 \text{ or } f^{*}_{\mu \nu}\,\mathrm{Im}\,\xi^{\nu}=0\)} (KKill);
\draw [->]
  (KConf) 
  -- node[above]{
    \begin{tabular}{c}
      $\mathrm{Im}\,\zeta^{\mu}=0$\\
      $\text{or } f^{*}_{\mu \nu}\,\mathrm{Im}\,\xi^{\nu}=0$
    \end{tabular}
  }
  (KKill);

% Box: Im Phi_ab Killing Yano tensor
\node (ImPhi) [draw, rectangle, text width=5.5cm, align=center, below=2.5cm of KillSpinor]
{\(\mathrm{Im}\,\Phi_{\mu \nu}\) Killing--Yano tensor};

% Arrows into ImPhi
\draw [->] (KillSpinor) -- node[left]{\(\mathrm{Im}\,\xi^{\mu}=0\)} (ImPhi);
\draw [->] (KKill) -- node[right]{\, \, \(\mathrm{Im}\,\alpha_{\mu}=0\)} (ImPhi);

% Box: Phi_ab complex Killing Yano tensor
\node (PhiComplex) [draw, rectangle, text width=5.5cm, align=center, below=2.5cm of ImPhi]
{\(\Phi_{\mu \nu}\) complex Killing--Yano tensor};

% Arrow from ImPhi down to PhiComplex
\draw [->] (ImPhi) -- node[left]{\(\mathrm{Re}\,\xi^{\mu}=0\)} (PhiComplex);

\end{tikzpicture}

    \caption{A flowchart derived in \cite{Dietz1980-dy} showing the classification of spacetimes admitting two independent null geodesic shear-free rays.}
    \label{fig:flowchart}
\end{figure}

Spacetimes admitting two geodesic shear-free rays were analyzed in \cite{Dietz1980-dy}, for which spacetimes admitting Killing-Yano tensors are contained within. The general family of spacetimes admitting two geodesic shear-free rays were also classified in that work. The authors also presented a flowchart showing the relations between the various structures, see figure \ref{fig:flowchart}. One could pick an explicit metric for these examples and follow the flowchart and methods of \cite{Dietz1980-dy} to solve for the Killing-Yano tensor. Here we will exploit that spacetimes in the Weyl double copy are known to include a Killing spinor, and then show that the bivector constructed from the Killing spinor in these cases is a Killing-Yano tensor.

To demonstrate that the spacetimes contain a Killing spinor, we can exploit the original proof for the splitting of the Weyl tensor from \cite{Walker:1970un,Hughston:1972qf}. In those works, it was demonstrated that if $o_{\lh{A}}$ and $\iota_{\lh{A}}$ are geodesic shear-free, and if an electromagnetic field of the form $f_{\lh{A}\lh{B}} = \mathcal{F} \, o_{(\lh{A}}\iota_{\lh{B})}$ exists satisfying Maxwell's vacuum equations, then the spinor:
\begin{equation}
    \chi_{\lh{A}\lh{B}} \equiv \mathcal{F}^{-3/2}f_{\lh{A}\lh{B}}, \label{chidef}
\end{equation}
is a Killing spinor, meaning:
\begin{equation}
    \nabla^{\rh{A}}_{(\lh{A}}\chi_{\lh{B}\lh{C})}=0.
\end{equation}
Importantly, the examples exhibiting the splitting in the sourced Weyl double copy all have a ``vacuum" electromagnetic spinor built from the principal null spinors of the Weyl spinor, that is to say they satisfy the conditions of the above theorem. According to figure \ref{fig:flowchart}, we need only show then that the quantity $\xi^\mu$ is real, which is defined in a similar way to \eqref{xidef}:
\begin{equation}
    \xi^\nu \equiv \frac{1}{3}\nabla_\mu \Phi^{\mu \nu}, \label{xipt2}
\end{equation}
where $\Phi_{\mu \nu}$ is defined the same way as \eqref{Phidef} with $\chi_{\lh{A}\lh{B}}$ as the Killing spinor. Having the imaginary part be zero is therefore equivalent to demonstrating:
\begin{equation}
    \nabla_{\mu} (\star \varphi^{\mu \nu}) \equiv i\nabla_{\mu} \left( \Phi^{\mu \nu} - \Bar{\Phi}^{\mu \nu} \right)  = 0,
\end{equation}
where:
\begin{equation}
    \varphi_{\mu \nu} \sigma^\mu_{\lh{A}\rh{A}}\sigma^\nu_{\lh{B}\rh{B}} \equiv \chi_{\lh{A}\lh{B}}\epsilon_{\rh{A}\rh{B}} + \Bar{\chi}_{\rh{A}\rh{B}}\epsilon_{\lh{A}\lh{B}}. \label{varphidef}
\end{equation}
So to summarize, the following steps can be taken to demonstrate that a spacetime admits a Killing-Yano tensor:
\begin{enumerate}
    \item The spacetime is Type D, with principal null directions geodesic and shear-free,
    \item There exists a vacuum electromagnetic spinor with shared principal null directions as the Weyl spinor. This defines a Killing spinor according to, \eqref{chidef}
    \item The divergence of the Hodge dual of the bivector formed from the Killing spinor according to \eqref{varphidef} must vanish.
\end{enumerate}
Note that if these conditions are satisfied, $\star \varphi_{\mu \nu}$ is a Killing-Yano tensor, and $\xi^\mu$ defined in \eqref{xipt2} is equivalent to \eqref{xidef}, and hence a Killing vector.

\subsubsection*{General non-accelerating black holes}

A general charged rotating non-accelerating black hole with cosmological constant and NUT charge, which we denote as Kerr-Newman-NUT-dS, is one such spacetime admitting a Killing-Yano tensor. It is a limit of the Plebanski-Demianski family of spacetimes, which are known to satisfy the first two requirements of admitting a Killing-Yano tensor \cite{Griffiths_Pleb}, namely the Weyl tensor is Type D with its principal null directions geodesic and shear-free, as well as the electromagnetic field sourcing the spacetime having aligned principal null directions with the Weyl tensor. Therefore, there automatically exists a Killing spinor formed from the vacuum electromagnetic spinor sourcing the spacetime via \eqref{chidef}.

Demonstrating the final condition takes some further calculations, for which we utilize the line element \cite{Griffiths_Pleb}:
\begin{equation}
    ds^2 = -\frac{Y(p,q)}{p^2+q^2}(d\tau-p^2 d\sigma)^2 +\frac{X(p,q)}{p^2+q^2}(d\tau+q^2 d\sigma)^2 + \frac{(p^2+q^2)}{X(p,q)}dp^2 + \frac{(p^2+q^2)}{Y(p,q)}dq^2,
\end{equation}
with:
\begin{equation}
    \begin{aligned}
        X(p,q) &\equiv \gamma -g^2+2 l p-\epsilon p^2 -\Lambda  p^4,\\
        Y(p,q) &\equiv \gamma +e^2-2 m q+\epsilon q^2-\Lambda  q^4, 
    \end{aligned}
\end{equation}
and the parameters $\{\gamma,e,g,l,\epsilon,\Lambda,m\}$ as arbitrary constants. Using the Newman-Penrose formalism \cite{newman1962approach}, we construct the vierbein via the null tetrad, which is induced from the principal spinors of the spacetimes:
\begin{equation}
    \begin{aligned}
        - o_{\lh{A}}o_{\rh{A}}\sigma_\mu^{\lh{A}\rh{A}}dx^\mu &= k_\mu dx^\mu = \frac{1}{\sqrt{2}}\sqrt{\frac{Y(p,q)}{(p^2+q^2)}}\left( -d\tau + p^2 d\sigma + \frac{p^2+q^2}{Y(p,q)}dq\right),\\
        - \iota_{\lh{A}}\iota_{\rh{A}}\sigma_\mu^{\lh{A}\rh{A}}dx^\mu &= \ell_\mu dx^\mu = \frac{1}{\sqrt{2}}\sqrt{\frac{Y(p,q)}{(p^2+q^2)}}\left( -d\tau + p^2 d\sigma - \frac{p^2+q^2}{Y(p,q)}dq\right),\\
        - o_{\lh{A}}\iota_{\rh{A}}\sigma_\mu^{\lh{A}\rh{A}}dx^\mu &= m_\mu dx^\mu = \frac{1}{\sqrt{2}}\sqrt{\frac{X(p,q)}{(p^2+q^2)}}\left( d\tau + q^2 d\sigma -i \frac{p^2+q^2}{X(p,q)}dp\right),\\
    \end{aligned}
\end{equation}
for which the Weyl tensor becomes:
\begin{equation}
    \Psi_{\lh{A}\lh{B}\lh{C}\lh{D}} = -6\,\frac{e^2+g^2-(l+i m) (p-i q)}{(p-i q) (p+i q)^3} o_{(\lh{A}}\iota_{\lh{B}}o_{\lh{C}}\iota_{\lh{D})},
\end{equation}
from which one can deduce the Maxwell vacuum spinor:
\begin{equation}
    f^{(0)}_{\lh{A}\lh{B}} \sim \frac{1}{(p+iq)^2} o_{(\lh{A}}\iota_{\lh{B})}.
\end{equation}
The Killing spinor can hence be identified via \eqref{chidef} as:
\begin{equation}
    \chi_{\lh{A}\lh{B}} \sim i(p+iq) o_{(\lh{A}}\iota_{\lh{B})},
\end{equation}
where the factor of $i$ is merely convention, which in turn induces a bivector from \eqref{varphidef}:
\begin{equation}
    \varphi_{\mu \nu} = \left(
\begin{array}{cccc}
 0 & 0 & p & -q \\
 0 & 0 & p q^2 & p^2 q \\
 -p & -p q^2 & 0 & 0 \\
 q & -p^2 q & 0 & 0 \\
\end{array}
\right),
\end{equation}
with its Hodge dual divergence free:
\begin{equation}
    \nabla_\mu (\star \varphi^{\mu \nu}) = 0.
\end{equation}

\subsubsection*{Spherical black hole metric expansion}

Consider a general stationary, spherically symmetric metric:
\begin{equation}
    ds^2 = -(1-f(r))dt^2 + \frac{1}{1-f(r)}dr^2 + r^2d \Omega^2,
\end{equation}
where the blackening function can be expanded in terms of powers of $1/r$ as:
\begin{equation}
    f(r) = \sum^m_{n=1}\frac{f_n}{r^n},
\end{equation}
with $m=1$ corresponding to the functional form of Schwarzschild, and $m=2$ for Reissner-Nordstrom. We can rewrite this metric into Kerr-Schild form via the transformation:
\begin{equation}
    dt \to dt + \frac{f(r)}{1-f(r)}dr,
\end{equation}
for which it transforms into:
\begin{equation}
    g_{\mu \nu} = \eta_{\mu \nu} + f(r)k_\mu k_\nu,
\end{equation}
with $k_\mu dx^\mu = -dt + dr$ and $\eta_{\mu \nu}$ being the Minkowski space metric in spherical coordinates. For Kerr-Schild metrics a necessary and sufficient condition for $k_\mu$ to be geodesic is:
\begin{equation}
    R_{\mu \nu}k^\mu k^\nu = 0, \label{geodKScond}
\end{equation}
and indeed, since $k^\mu \partial_\mu = \partial_t + \partial_r$ and the mixed-index Ricci tensor has the simple form:
\begin{equation}
    \tensor{R}{^\mu_\nu} = \frac{1}{2r^2}\left(
\begin{array}{cccc}
 r^2 f''(r)+2 r f'(r) & 0 & 0 & 0 \\
 0 & r^2 f''(r)+2 r f'(r) & 0 & 0 \\
 0 & 0 & 2 r f'(r)+2 f(r) & 0 \\
 0 & 0 & 0 & 2 r f'(r)+2 f(r) \\
\end{array}
\right),
\end{equation}
implies \eqref{geodKScond} is satisfied and hence $k_\mu$ is geodesic. The following vierbein can be utilized to convert to a Minkowski tetrad:
\begin{equation}
    \tensor{e}{^a_\mu} = \left(
\begin{array}{cccc}
 \sqrt{1-f(r)} & \frac{f(r)}{\sqrt{1-f(r)}} & 0 & 0 \\
 0 & 0 & r & 0 \\
 0 & 0 & 0 & r \sin (\theta ) \\
 0 & \frac{1}{\sqrt{1-f(r)}} & 0 & 0 \\
\end{array}
\right),
\end{equation}
which is a vierbein that automatically places the Weyl spinor in Type D form:
\begin{equation}
    \Psi_{\lh{A}\lh{B}\lh{C}\lh{D}} = -\frac{r^2 f''(r)-2 r f'(r)+2 f(r)}{2 r^2} o_{(\lh{A}}\iota_{\lh{B}}o_{\lh{C}}\iota_{\lh{D})}.
\end{equation}
It should be noted that \eqref{geodKScond} also implies that $k_\mu$ is a repeated principal null direction of the Weyl tensor, however we can also find the principal null directions via:
\begin{equation}
    o_{\lh{A}}o_{\rh{A}}\sigma_\mu^{\lh{A}\rh{A}} \propto (1,-1,0,0) \equiv k_\mu,
\end{equation}
\begin{equation}
    \iota_{\lh{A}}\iota_{\rh{A}}\sigma_\mu^{\lh{A}\rh{A}} \propto (1-f(r),1+f(r),0,0) \equiv \ell_\mu,
\end{equation}
for which both by construction are null. By defining the other complex null vector induced from the null tetrad:
\begin{equation}
    m_\mu \equiv o_{\lh{A}}\iota_{\rh{A}}\sigma_\mu^{\lh{A}\rh{A}},
\end{equation}
along with its complex conjugate, we can explicitly check to the geodesic and shear-free conditions:
\begin{equation}
    k^\mu m^\nu \nabla_\mu k_\nu = 0 = \ell^\mu \Bar{m}^\nu \nabla_\mu \ell_\nu,
\end{equation}
\begin{equation}
    m^\mu m^\nu \nabla_\mu k_\nu = 0 = \Bar{m}^\mu \Bar{m}^\nu \nabla_\mu \ell_\nu.
\end{equation}
Additionally, it has been noted \cite{Easson:2021asd} that the ``Schwarzschild" piece:
\begin{equation}
    f^{(0)}_{\lh{A}\lh{B}} \sim \frac{1}{r^2} o_{(\lh{A}}\iota_{\lh{B})},
\end{equation}
satisfies Maxwell's vacuum equations on this spacetime, and so, that is sufficient to demonstrate there exists a Killing spinor on the spacetime through \eqref{chidef}:
\begin{equation}
    \chi_{\lh{A}\lh{B}} = r o_{(\lh{A}}\iota_{\lh{B})}.
\end{equation}
This in turn induces a bivector from \eqref{varphidef}:
\begin{equation}
    \varphi_{\mu \nu} = \left(
\begin{array}{cccc}
 0 & r & 0 & 0 \\
 -r & 0 & 0 & 0 \\
 0 & 0 & 0 & 0 \\
 0 & 0 & 0 & 0 \\
\end{array}
\right),
\end{equation}
which in turn has its Hodge dual satisfy:
\begin{equation}
    \nabla_{\mu} (\star \varphi^{\mu \nu})  = 0,
\end{equation}
therefore demonstrating that there exists a Killing-Yano tensor in the spacetime, namely the Hodge dual itself:
\begin{equation}
    \star \varphi_{\mu \nu} = \left(
\begin{array}{cccc}
 0 & 0 & 0 & 0 \\
 0 & 0 & 0 & 0 \\
 0 & 0 & 0 & -r^3 \sin (\theta ) \\
 0 & 0 & r^3 \sin (\theta ) & 0 \\
\end{array}
\right),
\end{equation}
which can be readily shown to satisfy:
\begin{equation}
    \nabla_{(\alpha} \star \varphi_{\mu) \nu} = 0.
\end{equation}

\subsection*{Spacetimes without a Killing-Yano tensor}

These examples show how one can utilize the usual constructions of the Weyl double copy, such as the existence of a Killing spinor, to demonstrate a field strength built from a Killing vector as being equivalent to the Weyl double copy. This of course begs the question of how general this procedure is outside of Killing-Yano tensors, and two other examples can indeed be accounted for accordingly. The first are accelerating black holes included in the C-metrics, which no longer admit a Killing-Yano tensor, but are a subclass of the vacuum Plebanski-Demianski metric. Since they admit a Killing spinor, according to \cite{Dietz1980-dy} (see figure \ref{fig:flowchart}) the spacetime analogue is a \emph{conformal Killing-Yano tensor}, which itself has nice properties in vacuum. Namely, a Killing vector formed akin to \eqref{xidef} can be shown to exist in vacuum with similar structure \cite{Jezierski_2006}, which explains why the vacuum Plebanski-Demianski metrics directly correspond to their Kerr-Schild form \cite{Luna:2018dpt}. 

This naturally begs the question of charged C-metrics or generally sourced Plebanski-Demianski as analyzed in \cite{Easson:2022zoh}. They do exhibit two Killing vectors, however the structure of the Killing vectors are not linked in the same way when the spacetime is no longer vacuum. Since the sourced metrics observed in that work had single copies with magnetic charge (that could not be duality rotated away), this implies that they cannot have an understanding via a Killing vector, since Killing vectors necessarily have an unsourced Bianchi identity via \eqref{bianchi}. Type N gravitational waves \cite{Godazgar:2020zbv} additionally have been noted to have a Weyl double copy, and it should be noted that a Killing-Yano tensor is taken to be algebraically special (rather than algebraically general as was done here) the spacetime is necessarily Type N \cite{Dietz}, perhaps provided the needing connection. Finally, we note that the G\"odel double copy \cite{Kent2025} despite being neither Kerr-Schild nor Killing-Yano, is formed from the stationary Killing vector and so also admits some connection with the Weyl tensor.

\section{Discussion}

In this work we have undertaken an investigation into the root causes of the electromagnetic structures within, and the connections between, gravito-electromagnetism and the classical double copy. We found that prominent examples within these programs utilize properties of Killing vectors within these spacetimes to derive their electromagnetic structure. That is not to say all of their structure relies merely on a Killing vector, for example the ability to place the electromagnetic field on a flat background within the classical double copy relies on the Kerr-Schild form of the metric, not true for generic spacetimes with Killing vectors. Furthermore, the timelike Killing vector must also be geodesic to physically interpret the gravito-electromagnetic analogies drawn as tidal tensors. Nevertheless, the fact that Maxwell's equations are satisfied and that the tidal tensors built from said field strength are equivalent to gravitational tidal tensors is a consequence of Killing's equation. Furthermore, the Weyl double copy has primarily been related to the Kerr-Schild double copy via specific examples rather than general proofs. In this work we identify a class of spacetimes where this connection can be made concrete without identifying a specific metric, utilizing spacetimes admitting a Killing-Yano tensor. These spacetimes do not cover all spacetimes analyzed within the Weyl double copy but do contain a large class of physical solutions, and serve as an example for how a field strength built from a Killing vector can be directly related to the Weyl tensor.

This work leaves several routes open to further investigations on how Killing vectors play a role in known double copy constructions. Furthermore, the procedure outlined in \ref{sec2} can be utilized to generate new examples of classical double copies \cite{toappear}. We have identified several classes of spacetimes outside Killing-Yano type that should have a correspondence with Killing vectors, specifically vacuum spacetimes that contain Killing spinors whose tensorial analogue are conformal Killing-Yano tensors. These are natural candidates, since Killing spinors are central to the construction of many Weyl double copies, and they also define a Killing vector for vacuum spacetimes \cite{Jezierski_2006}. The zeroth copy in the Kerr-Schild double copy follows from the single copy equations on flat space, however the specific relation between the zeroth copy of the Kerr-Schild and Weyl double copies is another avenue of exploration.

Another route for further exploration is how this identification works on the level of the quantum field theory amplitudes double copy. At linearized level, twistors have been employed to relate the Weyl double copy in position space to the amplitudes double copy in momentum space \cite{Chacon:2021wbr,Chacon:2021lox,Luna:2022dxo}, and this can serve as a window into connecting these ideas at the level of amplitudes. It should also be noted that at least one example of the amplitudes double copy, between a bi-adjoint scalar theory and non-linear sigma model, has been explainable through the the group of isometries on the latter \cite{Cheung_2022}. Whether there is some understanding more generally if or how isometries play a part in the amplitudes double copy remains to be seen.

\section*{Acknowledgments}

It is a pleasure to thank Aaron Zimmerman for enlightening discussions. We also thank Gabriel Herczeg, Tucker Manton, Ricardo Monteiro, Donal O'Connell, and Andrew Svesko for comments and discussion regarding the manuscript.
Our work is supported by the National Science Foundation under grant 
No. PHY–2210562.
\bibliographystyle{jhep}
\bibliography{refs.bib}

\providecommand{\href}[2]{#2}\begingroup\raggedright\begin{thebibliography}{10}

\bibitem{Geroch1970-yf}
R.~Geroch, \emph{Multipole moments. {II}. curved space}, {\emph{J. Math. Phys.} {\bfseries 11} (1970) 2580}.

\bibitem{Hansen1974-xa}
R.~O. Hansen, \emph{Multipole moments of stationary space-times}, {\emph{J. Math. Phys.} {\bfseries 15} (1974) 46}.

\bibitem{Bel}
L.~Bel, \emph{Introduction d’un tenseur du quatrième ordre}, {\emph{C. R. Acad. Sci., Paris} {\bfseries 248} (1959) 1297}.

\bibitem{Penrose_Rindler_1984}
R.~Penrose and W.~Rindler, \emph{Spinors and Space-Time}, Cambridge Monographs on Mathematical Physics. Cambridge University Press, 1984.

\bibitem{Papapetrou}
A.~Papapetrou, \emph{Champs gravitationnels stationnaires \`a sym\'etrie axiale}, {\emph{Annales de l'institut Henri Poincar\'e. Section A, Physique Th\'eorique} {\bfseries 4} (1966) 83}.

\bibitem{Wald1974b}
R.~M. Wald, \emph{Black hole in a uniform magnetic field}, \href{https://doi.org/10.1103/PhysRevD.10.1680}{\emph{Phys. Rev. D} {\bfseries 10} (1974) 1680}.

\bibitem{Bern:2008qj}
Z.~Bern, J.~J.~M. Carrasco and H.~Johansson, \emph{{New Relations for Gauge-Theory Amplitudes}}, \href{https://doi.org/10.1103/PhysRevD.78.085011}{\emph{Phys. Rev. D} {\bfseries 78} (2008) 085011} [\href{https://arxiv.org/abs/0805.3993}{{\ttfamily 0805.3993}}].

\bibitem{Kawai:1985xq}
H.~Kawai, D.~C. Lewellen and S.~H.~H. Tye, \emph{{A Relation Between Tree Amplitudes of Closed and Open Strings}}, \href{https://doi.org/10.1016/0550-3213(86)90362-7}{\emph{Nucl. Phys. B} {\bfseries 269} (1986) 1}.

\bibitem{Bern:2013yya}
Z.~Bern, S.~Davies, T.~Dennen, Y.-t. Huang and J.~Nohle, \emph{{Color-Kinematics Duality for Pure Yang-Mills and Gravity at One and Two Loops}}, \href{https://doi.org/10.1103/PhysRevD.92.045041}{\emph{Phys. Rev.} {\bfseries D92} (2015) 045041} [\href{https://arxiv.org/abs/1303.6605}{{\ttfamily 1303.6605}}].

\bibitem{Monteiro:2014cda}
R.~Monteiro, D.~O'Connell and C.~D. White, \emph{{Black holes and the double copy}}, \href{https://doi.org/10.1007/JHEP12(2014)056}{\emph{JHEP} {\bfseries 12} (2014) 056} [\href{https://arxiv.org/abs/1410.0239}{{\ttfamily 1410.0239}}].

\bibitem{1965Kerr}
R.~P. {Kerr} and A.~{Schild}, \emph{{A new class of vacuum solutions of the Einstein field equations}},  in \emph{IV Centenario Della Nascita di Galileo Galilei, 1564-1964}, p.~222, Jan., 1965.

\bibitem{Luna:2015paa}
A.~Luna, R.~Monteiro, D.~O'Connell and C.~D. White, \emph{{The classical double copy for Taub\textendash{}NUT spacetime}}, \href{https://doi.org/10.1016/j.physletb.2015.09.021}{\emph{Phys. Lett. B} {\bfseries 750} (2015) 272} [\href{https://arxiv.org/abs/1507.01869}{{\ttfamily 1507.01869}}].

\bibitem{Chawla:2023bsu}
S.~Chawla and C.~Keeler, \emph{{Black hole horizons from the double copy}}, \href{https://doi.org/10.1088/1361-6382/acfe57}{\emph{Class. Quant. Grav.} {\bfseries 40} (2023) 225004} [\href{https://arxiv.org/abs/2306.02417}{{\ttfamily 2306.02417}}].

\bibitem{He:2023iew}
J.-L. He and J.-H. Huang, \emph{{Cosmological horizons from classical double copy}}, \href{https://doi.org/10.1016/j.physletb.2024.138579}{\emph{Phys. Lett. B} {\bfseries 851} (2024) 138579} [\href{https://arxiv.org/abs/2312.00972}{{\ttfamily 2312.00972}}].

\bibitem{Bahjat-Abbas:2020cyb}
N.~Bahjat-Abbas, R.~Stark-Much{\~a}o and C.~D. White, \emph{{Monopoles, shockwaves and the classical double copy}}, \href{https://doi.org/10.1007/JHEP04(2020)102}{\emph{JHEP} {\bfseries 04} (2020) 102} [\href{https://arxiv.org/abs/2001.09918}{{\ttfamily 2001.09918}}].

\bibitem{Godazgar:2020zbv}
H.~Godazgar, M.~Godazgar, R.~Monteiro, D.~Peinador~Veiga and C.~N. Pope, \emph{{Weyl Double Copy for Gravitational Waves}}, \href{https://doi.org/10.1103/PhysRevLett.126.101103}{\emph{Phys. Rev. Lett.} {\bfseries 126} (2021) 101103} [\href{https://arxiv.org/abs/2010.02925}{{\ttfamily 2010.02925}}].

\bibitem{CarrilloGonzalez:2022mxx}
M.~Carrillo~Gonz\'alez, A.~Momeni and J.~Rumbutis, \emph{{Cotton double copy for gravitational waves}}, \href{https://doi.org/10.1103/PhysRevD.106.025006}{\emph{Phys. Rev. D} {\bfseries 106} (2022) 025006} [\href{https://arxiv.org/abs/2202.10476}{{\ttfamily 2202.10476}}].

\bibitem{Andrzejewski:2019hub}
K.~Andrzejewski and S.~Prencel, \emph{{From polarized gravitational waves to analytically solvable electromagnetic beams}}, \href{https://doi.org/10.1103/PhysRevD.100.045006}{\emph{Phys. Rev. D} {\bfseries 100} (2019) 045006} [\href{https://arxiv.org/abs/1901.05255}{{\ttfamily 1901.05255}}].

\bibitem{Ilderton:2018lsf}
A.~Ilderton, \emph{{Screw-symmetric gravitational waves: a double copy of the vortex}}, \href{https://doi.org/10.1016/j.physletb.2018.04.069}{\emph{Phys. Lett. B} {\bfseries 782} (2018) 22} [\href{https://arxiv.org/abs/1804.07290}{{\ttfamily 1804.07290}}].

\bibitem{alawadhi2020s}
R.~Alawadhi, D.~S. Berman, B.~Spence and D.~P. Veiga, \emph{S-duality and the double copy}, {\emph{Journal of High Energy Physics} {\bfseries 2020} (2020) 1}.

\bibitem{Luna:2018dpt}
A.~Luna, R.~Monteiro, I.~Nicholson and D.~O'Connell, \emph{{Type D Spacetimes and the Weyl Double Copy}}, \href{https://doi.org/10.1088/1361-6382/ab03e6}{\emph{Class. Quant. Grav.} {\bfseries 36} (2019) 065003} [\href{https://arxiv.org/abs/1810.08183}{{\ttfamily 1810.08183}}].

\bibitem{Chacon:2021wbr}
E.~Chac\'on, S.~Nagy and C.~D. White, \emph{{The Weyl double copy from twistor space}}, \href{https://doi.org/10.1007/JHEP05(2021)239}{\emph{JHEP} {\bfseries 05} (2021) 2239} [\href{https://arxiv.org/abs/2103.16441}{{\ttfamily 2103.16441}}].

\bibitem{Chacon:2021lox}
E.~Chac\'on, S.~Nagy and C.~D. White, \emph{{Alternative formulations of the twistor double copy}}, \href{https://doi.org/10.1007/JHEP03(2022)180}{\emph{JHEP} {\bfseries 03} (2022) 180} [\href{https://arxiv.org/abs/2112.06764}{{\ttfamily 2112.06764}}].

\bibitem{Luna:2022dxo}
A.~Luna, N.~Moynihan and C.~D. White, \emph{{Why is the Weyl double copy local in position space?}}, \href{https://doi.org/10.1007/JHEP12(2022)046}{\emph{JHEP} {\bfseries 12} (2022) 046} [\href{https://arxiv.org/abs/2208.08548}{{\ttfamily 2208.08548}}].

\bibitem{Easson:2023dbk}
D.~A. Easson, G.~Herczeg, T.~Manton and M.~Pezzelle, \emph{{Isometries and the double copy}}, \href{https://doi.org/10.1007/JHEP09(2023)162}{\emph{JHEP} {\bfseries 09} (2023) 162} [\href{https://arxiv.org/abs/2306.13687}{{\ttfamily 2306.13687}}].

\bibitem{Carrillo-Gonzalez:2017iyj}
M.~Carrillo-Gonz\'alez, R.~Penco and M.~Trodden, \emph{{The classical double copy in maximally symmetric spacetimes}}, \href{https://doi.org/10.1007/JHEP04(2018)028}{\emph{JHEP} {\bfseries 04} (2018) 028} [\href{https://arxiv.org/abs/1711.01296}{{\ttfamily 1711.01296}}].

\bibitem{Bahjat-Abbas:2017htu}
N.~Bahjat-Abbas, A.~Luna and C.~D. White, \emph{{The Kerr-Schild double copy in curved spacetime}}, \href{https://doi.org/10.1007/JHEP12(2017)004}{\emph{JHEP} {\bfseries 12} (2017) 004} [\href{https://arxiv.org/abs/1710.01953}{{\ttfamily 1710.01953}}].

\bibitem{Alka__2021}
G.~Alkaç, M.~K. Gümüş and M.~Tek, \emph{The kerr-schild double copy in lifshitz spacetime}, \href{https://doi.org/10.1007/jhep05(2021)214}{\emph{Journal of High Energy Physics} {\bfseries 2021} (2021) }.

\bibitem{Alkac:2021bav}
G.~Alkac, M.~K. Gumus and M.~Tek, \emph{{The Classical Double Copy in Curved Spacetime}},  \href{https://arxiv.org/abs/2103.06986}{{\ttfamily 2103.06986}}.

\bibitem{Costa_2008}
L.~F.~O. Costa and C.~A.~R. Herdeiro, \emph{Gravitoelectromagnetic analogy based on tidal tensors}, \href{https://doi.org/10.1103/physrevd.78.024021}{\emph{Physical Review D} {\bfseries 78} (2008) }.

\bibitem{Costa2014}
L.~F.~O. Costa and J.~Nat{\'a}rio, \emph{Gravito-electromagnetic analogies}, \href{https://doi.org/10.1007/s10714-014-1792-1}{\emph{General Relativity and Gravitation} {\bfseries 46} (2014) 1792}.

\bibitem{Walker:1970un}
M.~Walker and R.~Penrose, \emph{{On quadratic first integrals of the geodesic equations for type [22] spacetimes}}, \href{https://doi.org/10.1007/BF01649445}{\emph{Commun. Math. Phys.} {\bfseries 18} (1970) 265}.

\bibitem{Hughston:1972qf}
L.~P. Hughston, R.~Penrose, P.~Sommers and M.~Walker, \emph{{On a quadratic first integral for the charged particle orbits in the charged kerr solution}}, \href{https://doi.org/10.1007/BF01645517}{\emph{Commun. Math. Phys.} {\bfseries 27} (1972) 303}.

\bibitem{Easson:2021asd}
D.~A. Easson, T.~Manton and A.~Svesko, \emph{{Sources in the Weyl Double Copy}}, \href{https://doi.org/10.1103/PhysRevLett.127.271101}{\emph{Phys. Rev. Lett.} {\bfseries 127} (2021) 271101} [\href{https://arxiv.org/abs/2110.02293}{{\ttfamily 2110.02293}}].

\bibitem{Easson:2022zoh}
D.~A. Easson, T.~Manton and A.~Svesko, \emph{{Einstein-Maxwell theory and the Weyl double copy}}, \href{https://doi.org/10.1103/PhysRevD.107.044063}{\emph{Phys. Rev. D} {\bfseries 107} (2023) 044063} [\href{https://arxiv.org/abs/2210.16339}{{\ttfamily 2210.16339}}].

\bibitem{alkac2024regularizedweyldoublecopy}
G.~Alkac, M.~K. Gumus, O.~Kasikci, M.~A. Olpak and M.~Tek, \emph{Regularized weyl double copy},  2024.

\bibitem{Armstrong-Williams:2024bog}
K.~Armstrong-Williams, N.~Moynihan and C.~D. White, \emph{{Deriving Weyl double copies with sources}},  \href{https://arxiv.org/abs/2407.18107}{{\ttfamily 2407.18107}}.

\bibitem{Kent2025}
B.~Kent, T.~Manton and S.~Shashi, \emph{Background ambiguity and the g{\"o}del double copy}, \href{https://doi.org/10.1007/JHEP03(2025)033}{\emph{Journal of High Energy Physics} {\bfseries 2025} (2025) 33} [\href{https://arxiv.org/abs/2411.04207}{{\ttfamily 2411.04207}}].

\bibitem{Stephani:2003tm}
H.~Stephani, D.~Kramer, M.~A.~H. MacCallum, C.~Hoenselaers and E.~Herlt, \emph{{Exact solutions of Einstein's field equations}}, Cambridge Monographs on Mathematical Physics. Cambridge Univ. Press, Cambridge, 2003, \href{https://doi.org/10.1017/CBO9780511535185}{10.1017/CBO9780511535185}.

\bibitem{Wald:1984rg}
R.~M. Wald, \emph{{General Relativity}}. Chicago Univ. Pr., Chicago, USA, 1984, \href{https://doi.org/10.7208/chicago/9780226870373.001.0001}{10.7208/chicago/9780226870373.001.0001}.

\bibitem{1969JMP....10.1842D}
G.~C. {Debney}, R.~P. {Kerr} and A.~{Schild}, \emph{{Solutions of the Einstein and Einstein-Maxwell Equations}}, \href{https://doi.org/10.1063/1.1664769}{\emph{Journal of Mathematical Physics} {\bfseries 10} (1969) 1842}.

\bibitem{Ortaggio_2024}
M.~Ortaggio and A.~Srinivasan, \emph{Charging kerr-schild spacetimes in higher dimensions}, \href{https://doi.org/10.1103/physrevd.110.044035}{\emph{Physical Review D} {\bfseries 110} (2024) }.

\bibitem{Ortaggio_2024b}
M.~Ortaggio, V.~Pravda and A.~Pravdová, \emph{Kerr-schild double copy for kundt spacetimes of any dimension}, \href{https://doi.org/10.1007/jhep02(2024)069}{\emph{Journal of High Energy Physics} {\bfseries 2024} (2024) }.

\bibitem{Ett_2010}
B.~Ett and D.~Kastor, \emph{An extended kerr–schild ansatz}, \href{https://doi.org/10.1088/0264-9381/27/18/185024}{\emph{Classical and Quantum Gravity} {\bfseries 27} (2010) 185024}.

\bibitem{Dixon1973-xj}
W.~G. Dixon, \emph{The definition of multipole moments for extended bodies}, {\emph{Gen. Relativ. Gravit.} {\bfseries 4} (1973) 199}.

\bibitem{Mathisson2010-tf}
M.~Mathisson, \emph{Republication of: New mechanics of material systems}, {\emph{Gen. Relativ. Gravit.} {\bfseries 42} (2010) 1011}.

\bibitem{Costa_2016}
L.~F.~O. Costa, J.~Natário and M.~Zilhão, \emph{Spacetime dynamics of spinning particles: Exact electromagnetic analogies}, \href{https://doi.org/10.1103/physrevd.93.104006}{\emph{Physical Review D} {\bfseries 93} (2016) }.

\bibitem{Huggett:1986fs}
S.~Huggett and K.~Tod, \emph{{AN INTRODUCTION TO TWISTOR THEORY}}. London Mathematical Society Student Texts, 9, 1986.

\bibitem{Godel:1949ga}
K.~Godel, \emph{{An Example of a new type of cosmological solutions of Einstein's field equations of graviation}}, \href{https://doi.org/10.1103/RevModPhys.21.447}{\emph{Rev. Mod. Phys.} {\bfseries 21} (1949) 447}.

\bibitem{Som}
M.~M. Som and A.~K. Raychaudhuri, \emph{Cylindrically symmetric charged dust distributions in rigid rotation in general relativity}, {\emph{Proceedings of the Royal Society of London. Series A, Mathematical and Physical Sciences} {\bfseries 304} (1968) 81}.

\bibitem{lense-thirring}
B.~{Mashhoon}, F.~W. {Hehl} and D.~S. {Theiss}, \emph{{On the gravitational effects of rotating masses: the Thirring-Lense papers.}}, \href{https://doi.org/10.1007/BF00762913}{\emph{General Relativity and Gravitation} {\bfseries 16} (1984) 711}.

\bibitem{Costa_2009}
L.~F.~O. Costa and C.~A.~R. Herdeiro, \emph{Reference frames and the physical gravito-electromagnetic analogy}, \href{https://doi.org/10.1017/S1743921309990111}{\emph{Proceedings of the International Astronomical Union} {\bfseries 5} (2009) 31–39}.

\bibitem{plebanski1976rotating}
J.~F. Plebanski and M.~Demianski, \emph{Rotating, charged, and uniformly accelerating mass in general relativity}, {\emph{Annals of Physics} {\bfseries 98} (1976) 98}.

\bibitem{Chawla:2022ogv}
S.~Chawla and C.~Keeler, \emph{{Aligned Fields Double Copy to Kerr-NUT-(A)dS}},  \href{https://arxiv.org/abs/2209.09275}{{\ttfamily 2209.09275}}.

\bibitem{Dietz}
W.~Dietz and R.~Rüdiger, \emph{Space–times admitting killing–yano tensors. i}, \href{https://doi.org/10.1098/rspa.1981.0056}{\emph{Proceedings of the Royal Society of London. A. Mathematical and Physical Sciences} {\bfseries 375} (1981) 361–378}.

\bibitem{Griffiths_Pleb}
J.~B. GRIFFITHS and J.~PODOLSK\'{Y}, \emph{A new look at the plebaŃski–demiaŃski family of solutions}, \href{https://doi.org/10.1142/S0218271806007742}{\emph{International Journal of Modern Physics D} {\bfseries 15} (2006) 335} [\href{https://arxiv.org/abs/https://doi.org/10.1142/S0218271806007742}{{\ttfamily https://doi.org/10.1142/S0218271806007742}}].

\bibitem{Dietz1980-dy}
W.~Dietz and R.~Rudiger, \emph{Shearfree congruences of null geodesies and killing tensors}, {\emph{Gen. Relativ. Gravit.} {\bfseries 12} (1980) 545}.

\bibitem{newman1962approach}
E.~Newman and R.~Penrose, \emph{An approach to gravitational radiation by a method of spin coefficients}, {\emph{Journal of Mathematical Physics} {\bfseries 3} (1962) 566}.

\bibitem{Jezierski_2006}
J.~Jezierski and M.~Łukasik, \emph{Conformal yano–killing tensor for the kerr metric and conserved quantities}, \href{https://doi.org/10.1088/0264-9381/23/9/008}{\emph{Classical and Quantum Gravity} {\bfseries 23} (2006) 2895–2918}.

\bibitem{toappear}
B.~Kent and A.~Zimmerman, \emph{In preparation},  2025.

\bibitem{Cheung_2022}
C.~Cheung, J.~Parra-Martinez and A.~Sivaramakrishnan, \emph{On-shell correlators and color-kinematics duality in curved symmetric spacetimes}, \href{https://doi.org/10.1007/jhep05(2022)027}{\emph{Journal of High Energy Physics} {\bfseries 2022} (2022) }.

\end{thebibliography}\endgroup

\end{document}